\documentclass[11pt,prd,preprintnumbers,amsmath,amssymb, superscriptaddress]{revtex4}
\pdfoutput=1
\usepackage{amssymb}
\usepackage{amsmath}
\usepackage{enumerate}
\usepackage{graphicx}
\usepackage{subfigure}
\usepackage{multirow}
\usepackage{hyperref}
\usepackage{url}

\def\be{\begin{equation}}
\def\ee{\end{equation}}
\def\ba{\begin{array}}
\def\ea{\end{array}}
\makeatletter

\def\be{\begin{equation}}
\def\ee{\end{equation}}
\def\ba{\begin{eqnarray}}
\def\ea{\end{eqnarray}}
\def\beas{\begin{eqnarray*}}
\def\eeas{\end{eqnarray*}}
\def\sla{\raise.15ex\hbox{$/$}\kern-.57em}

\newcommand{\BSM}{\text{BSM}}
\newcommand{\MET}{\text{MET}}
\newcommand{\SM}{\text{SM}}
\newcommand{\CMS}{\text{CMS}}

\newcommand{\NLSP}{\text{NLSP}}

\newcommand{\CLs}{{\text{ CL}_s}}

\newcommand{\GeV}{\text{ GeV}}
\newcommand{\TeV}{\text{ TeV}}

\newcommand{\fb}{\text{ fb}}

\begin{document}
\title{\Large\bf Excess Higgs Production in Neutralino Decays}

\author{Kiel Howe}

\author{Prashant Saraswat}

\affiliation{Stanford Institute for Theoretical Physics, Department of Physics, Stanford University, Stanford, CA 94305}

\preprint{SU-ITP-12/23}


\begin{abstract}
The ATLAS and CMS experiments have recently claimed discovery of a Higgs boson-like particle at $\sim 5\sigma$ confidence and are beginning to test the Standard Model predictions for its production and decay. In a variety of supersymmetric models, a neutralino NLSP can decay dominantly to the Higgs and the LSP. In natural SUSY models, a light third generation squark decaying through this chain can lead to large excess Higgs production while evading existing BSM searches. Such models can be observed at the 8 TeV LHC in channels exploiting the rare diphoton decays of the Higgs produced in the cascade decay. Identifying a diphoton resonance in association with missing energy, a lepton, or $b$-tagged jets is a promising search strategy for discovery of these models, and would immediately signal new physics involving production of a Higgs boson. We also discuss the possibility that excess Higgs production in these SUSY decays can be responsible for enhancements of up to 50\% over the SM prediction for the observed rate in the existing inclusive diphoton searches, a scenario which would likely by the end of the 8 TeV run be accompanied by excesses in the $\gamma\gamma +\ell/\MET$ and SUSY multi-lepton/$b$ searches and a potential discovery in a $\gamma\gamma+2b$ search.
\end{abstract}

\maketitle

\pagebreak

\tableofcontents

\pagebreak

\section{Introduction} \label{sec:Intro}

The ATLAS and CMS experiments at the Large Hadron Collider (LHC) have recently reported discovery of a new particle consistent with the Higgs boson of the Standard Model~\cite{ATLASDiscovery, CMSDiscovery}. Each experiment independently reports nearly $5\sigma$ combined local significance for a mass of $m_h\sim 125 \GeV$, largely fueled by the searches for resonances in diphoton and $ZZ^* \to 4l$ final states. 

The discovery of the Higgs marks the final chapter in the quest to discover the particle content of the Standard Model. It also represents the start of a new era of Higgs physics at the LHC. With additional data and further analyses the experiments will be able to study the properties of this new particle, in particular its production modes and decay branching ratios. The Standard Model makes very definite predictions for these properties (up to calculational uncertainties) given the mass of the Higgs, so any significant deviation from the expected values would be an immediate sign of new physics. 

There are many ways that physics beyond the Standard Model (BSM) can manifest itself in Higgs events. One possibility is that the Higgs couplings to known SM particles are modified by new physics, such as loop effects of new particles, mixing of the Higgs with other new states (extended Higgs sectors), and Higgs compositeness. Such effects will modify both the decay branching ratios and production modes of the Higgs, potentially giving observable deviations from SM predictions. Several works perform general analyses of these possibilities in the context of recent LHC Higgs results  \cite{Azatov:2012bz,Benbrik:2012rm,Carmi:2012in,Ellis:2012rx,Espinosa:2012im,Giardino:2012ww,Giardino:2012dp,Klute:2012pu,Low:2012rj}.

A second possibility is that on-shell production of BSM particles can lead to Higgs production in the decay cascade. This possibility has been discussed as a discovery mode for an SM-like Higgs boson in several models~\cite{delAguila:1989ba,delAguila:1989rq,Sher:1999ae,AguilarSaavedra:2006gw,Kilic:2010fs,Carmona:2012jk,Kribs:2009yh,Bandyopadhyay:2011qm,Son:2012mb}, for the extra Higgs states of the minimal supersymmetric standard model (MSSM) and its extensions~\cite{Bandyopadhyay:2010tv,Datta:2001qs,Datta:2003iz,deCampos:2008ic,Fowler:2009ay,Gori:2011hj,Kribs:2010hp,Stal:2011cz}, and for composite Higgs models \cite{Azuelos:2004dm, Kribs:2010ii, Azatov:2012rj,Vignaroli:2012sf}. Conversely, the discovery sensitivity for BSM physics through Higgs production has been studied in the context of the MSSM with a bino LSP~\cite{Byakti:2012qk,Ghosh:2012mc,Baer:2012ts}, a gravitino LSP~\cite{Asano:2010ut,Kats:2011qh,Matchev:1999ft,ChiAtTeV,ChiAtLHC}, and additional feebly coupled LSPs \cite{PhotiniShadow,Thaler:2011me,Das:2012rr}, as well as in the context of vector-like heavy quarks \cite{AguilarSaavedra:2006gv, AguilarSaavedra:2009es, Harigaya:2012ir, Girdhar:2012vn,Vignaroli:2012nf, Azatov:2012rj}.
Higgs production in cascade decays has also been studied as a way to probe the details of the the spectra and couplings of BSM particles after their discovery~\cite{Baer:1992ef,Bandyopadhyay:2008fp,Bandyopadhyay:2008sd,Bhattacherjee:2011vz,Hinchliffe:1996iu,Huitu:2008sa}. 

In this work we consider a class of SUSY models with neutralino NLSPs which dominantly decay to a Higgs boson. In particular we focus on models with light third generation squarks, which are motivated by naturalness of the electroweak scale. Such models can remain unconstrained by existing SUSY searches while still having total SUSY production cross sections at the 7 or 8 TeV LHC of up to a few picobarns-- comparable to the total SM Higgs production cross section and significantly larger than the cross section for associated production in the SM. The sensitivity of SUSY searches is limited due to the suppression of missing energy when a light NLSP ($\lesssim 200 \GeV$) decays to a Higgs and the LSP, as has been explored previously in more general models~\cite{PhotiniShadow,Kats:2011qh,Das:2012rr}.

The dominant decays for a light Higgs are $h\to b\overline{b}$ and $h\to WW^*$, and much of the existing work on BSM Higgs production has focused on these modes. The $h\to b\bar{b}$ mode suggests search channels with many $b$'s or searches for `Higgs-tagged' boosted jets~\cite{Kribs:2009yh,Kribs:2010hp,Kribs:2010ii,Gori:2011hj,Byakti:2012qk}, while the $h\to WW^*$ decay suggests searches in leptonic channels with missing energy. The many-$b$ modes have large QCD backgrounds when the MET signature is suppressed, while the leptonic modes have smaller branching fractions and can have significant backgrounds as well. Furthermore, it can be difficult to link a new physics signal with the Higgs in these channels, especially in the presence of other hard objects from the SUSY cascade. 

We will focus instead on the potential for observing such models using the rare but distinctive diphoton decay mode of the Higgs, which has a branching fraction of 0.22\% for a 125 GeV SM Higgs.  We show that searches requiring a diphoton with invariant mass near 125 GeV plus other objects such as $b$-tagged jets, missing energy or leptons can target the models we consider while having very low SM backgrounds (a similar point is made in the context of searching for heavy vector-like quarks in ref.~\cite{Azatov:2012rj}). Since such searches look for a narrow resonance in the diphoton mass, the backgrounds can be directly determined from data and the observation of a peak near 125 GeV would strongly suggest new physics involving production of an on-shell Higgs. We find that in the $8$ TeV LHC run the $\gamma\gamma+2b$ channel could give a strong signal in regions of parameter space where other searches are not sensitive, and that the $\gamma\gamma+\MET$ and $\gamma\gamma+\ell$ channels have sensitivity comparable to and in some regions exceeding that of existing SUSY searches. These results are summarized in fig.~\ref{fig:CombinedLimits}. 

Although targeted $\gamma \gamma + X$ searches have the best potential to probe our models, for sufficiently light squarks even inclusive searches for the SM Higgs boson in the resonant $\gamma \gamma$ and $ZZ^* \to 4\ell$ channels can have some sensitivity. Enhancements of 20-50\% over the SM rates can be accommodated within current bounds from other searches. Searches in non-resonant channels, such as $h \to WW^*$, $h \to \tau \tau$ and $Vh \to V b \bar{b}$, generally include additional cuts that tend to reject SUSY events for the models we consider, so no excess rate is predicted for these analyses. Our models therefore predict a very particular pattern of excesses in certain Higgs searches which can be observable with sufficient data. Such excesses can be consistent with the currently observed rates in the ATLAS and CMS Higgs searches, which report a higher than expected rate in the diphoton channels. We find that in the regions of parameter space where the excess rate in SM Higgs searches is significant, strong signals are expected in the  $\gamma\gamma + b/\MET$ modes within the 8 TeV run. 

We now give an outline of the rest of this work. In sec.~\ref{sec:Models} we describe models in which a neutralino NLSP can decay dominantly to the Higgs and in which the third generation squarks are light and give the dominant SUSY production cross section. In sec.~\ref{sec:Constraints} we consider the limits on these models from existing LHC SUSY searches.  We then consider the signal in the Higgs to diphoton channels (sec.~\ref{sec:Diphoton}), determining the sensitivity of existing searches for $\gamma\gamma+\ell$ and $\gamma\gamma+\MET$, and suggesting a new resonant search for diphotons in association with a $b$-tagged jet. The possibility that current hints of excesses in the inclusive Higgs diphoton channel can be explained within this model is discussed in sec.~\ref{sec:OtherSearch}. We  also briefly discuss the signals from direct electroweak production of neutralinos (sec.~\ref{sec:DirectProd}). We then conclude in sec.~\ref{sec:Conclusions} with an outlook for the rest of the 8 TeV run and a discussion of more general motivations for searches for new physics in the Higgs + $X$ channel.

\section{Models}\label{sec:Models}

In this section, we describe a set of models in which neutralino NLSPs decay dominantly to the LSP and a Higgs. An important consequence of this mechanism for Higgs production is that if the LSP is light and the splitting between the NLSP and the Higgs is not too large, the MET signature in the SUSY cascade is suppressed. This can relax the collider constraints on superpartners, as we will discuss in detail for our model in sec.~\ref{sec:Constraints} and is explored for gluino and degenerate squark simplified models and the c(N)MSSM in refs.~\cite{PhotiniShadow,Das:2012rr,Kats:2011qh}. In particular, this can allow for large colored production cross sections with only weak constraints from existing searches.

Amongst the SUSY particles the squarks and gluinos typically have the largest production cross sections due to their color charge. If the squarks are degenerate in mass, existing SUSY searches strongly constrain their mass in the simplest $R$-parity conserving scenarios. Squarks decaying directly to light neutralinos must be heavier than $\sim$TeV~\cite{ATLASjetsMET5, CMSjetsMET5}. These bounds are in tension with naturalness of the electroweak scale in SUSY, as if the stops are heavy then they give large corrections to the Higgs masses that must be tuned away.

It is possible however to recover SUSY naturalness if only the third generation squarks are light, with masses $\sim$ few hundred GeV, with the other squarks much heavier-- so-called ``natural spectra"~\cite{Dimopoulos:1995mi,Pomarol:1995xc,Cohen:1996vb,Agashe:1998zz,Kaplan:1998jk,Aharony:2010ch,Barbieri:2010ar,Craig:2011yk,Larsen:2012rq,Craig:2012hc,Craig:2012di,Baer:2012uy,Cohen:2012rm,Randall:2012dm}.
 (Some extension of the MSSM is then necessary if the lightest Higgs is to have a mass near 125 GeV; a number of possibilities have been explored~\cite{Dermisek:2009si, Graham:2009gy, Martin:2009bg, Choi:2005hd, Kitano:2005wc,  Chacko:2005ra, Ellis:1988er, Espinosa:1998re, Batra:2003nj, Maloney:2004rc, Casas:2003jx, Brignole:2003cm, Harnik:2003rs, Chang:2004db,  Delgado:2005fq,  Birkedal:2004zx, Babu:2004xg, Choi:2006xb}.) This scenario is much less constrained than the case of degenerate squarks since the third-generation squarks have smaller production cross sections and can produce less missing energy in their decays \cite{Brust:2011tb,Papucci:2011wy,Lee:2012sy}.

Combining these two elements, a natural squark spectrum and a neutralino NLSP decaying dominantly to the Higgs, we find a set of models that evade constraints from existing searches but give large cross sections for Higgs production, potentially giving striking signatures in searches exploiting the Higgs resonance. The key features of these realistic models can be cast into a simplified model (section~\ref{sec:SimpModel}) which will be our main focus for the rest of this work.

\subsection{NLSP Decays to the Higgs}\label{sec:ChiToHiggs}

If $R$-parity is conserved, the two-body decays of a neutralino NLSP to a neutral LSP can produce an on- or off-shell Higgs, $Z$, or photon. In this section we describe some types of models in which a wino, bino, or higgsino NLSP decays dominantly to an on-shell Higgs and a neutral LSP, making this channel a promising way to probe such a model at the LHC. 

\begin{figure}
		  \centering
        \subfigure[]{
                \includegraphics[width=2in]{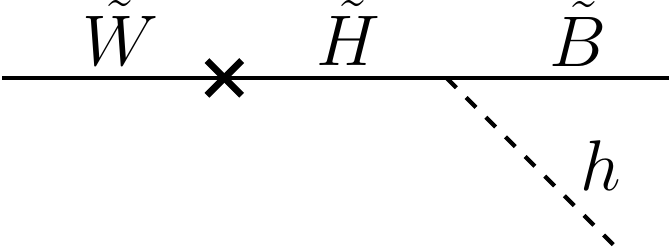}
                \label{fig:Wtoh}
		  }
        \hspace{5mm} 
        \subfigure[]{
                \includegraphics[width=2in]{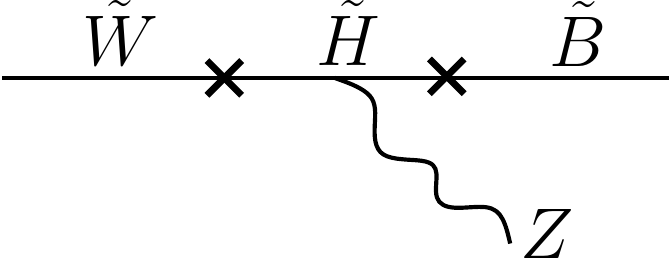}
                \label{fig:WtoZ}
		  }
        \caption{Diagrams for decays of a wino NLSP to a bino LSP and a Higgs or a $Z$. The neutralinos are shown in the gauge eigenbasis, with mixing represented by interaction vertices (crosses). The decay of the wino to the bino and a Higgs proceeds through mixing of one of the states with the Higgsino, while the decay to the $Z$ requires at least two mixings and is further suppressed.}\label{fig:MixingFeyn}
\end{figure}

\begin{figure}
		  \centering
        \subfigure[]{
                \includegraphics[width=3in]{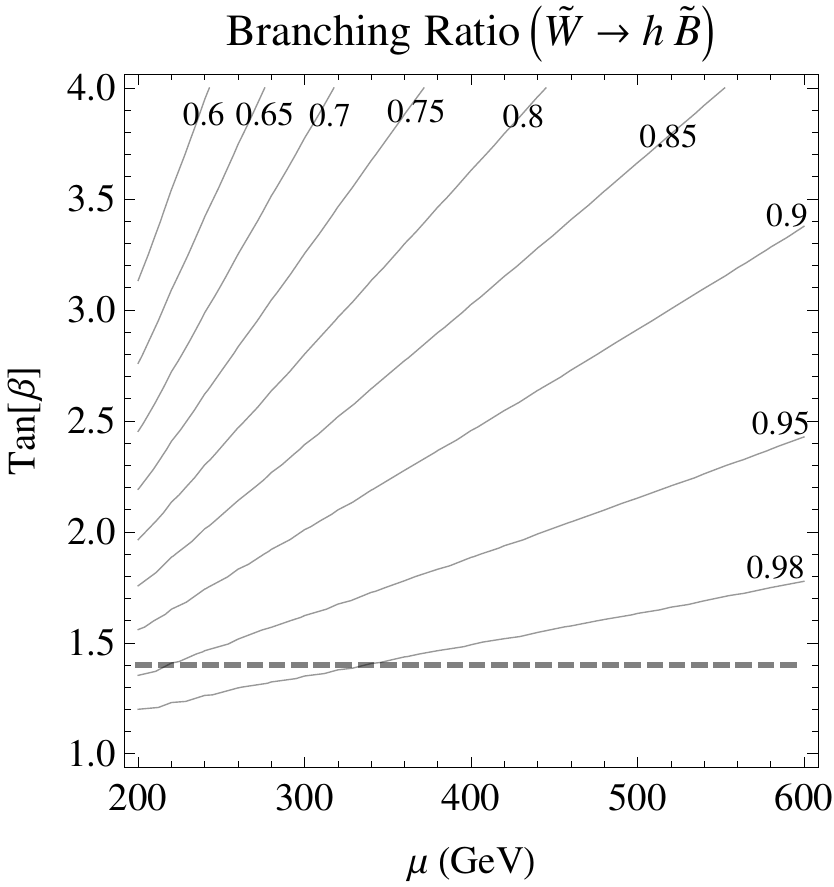}
                \label{fig:BRWtoh}
		  }
        \hspace{5mm} 
        \subfigure[]{
                \includegraphics[width=3in]{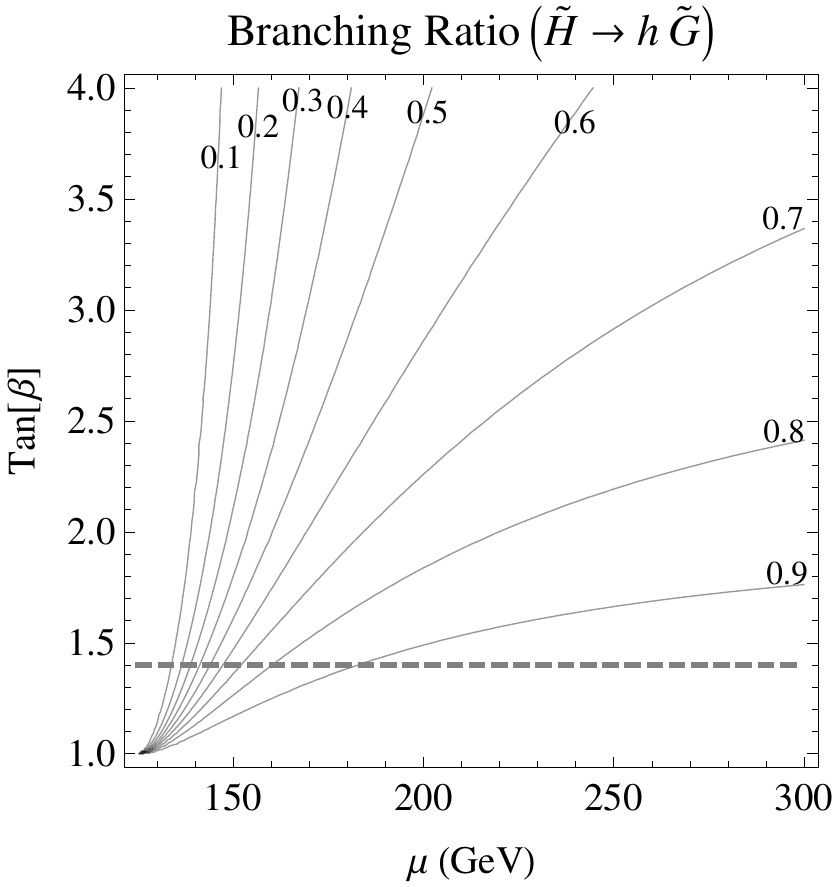}
                \label{fig:BRHtoh}
		  }
        \caption{(a): Branching ratio of a neutral wino NLSP to a Higgs and a bino LSP as a function of $\mu$ and $\tan \beta$, for a wino mass of 175 GeV and a Higgs of mass 125 GeV in the decoupling limit. The branching ratio is computed using tree-level neutralino masses and mixings, with the input $M_1$ = 0 and choosing $M_2$ such that the wino mass is 175 GeV. The resulting bino mass varies throughout this parameter space but remains very light ($< 10 \GeV$). (b): Branching ratio of a higgsino NLSP to a Higgs and a bino LSP as a function of higgsino mass and $\tan \beta$, for large $M_1$, $M_2$, a light gravitino, and a Higgs of mass 125 GeV in the decoupling limit, using the formulae of~\cite{ChiAtTeV}. This plot assumes $\text{sign}(\mu(\frac{M_1}{M_2}+\tan^2\theta_W))<0$, which causes the higgsino-$Z$-gravitino coupling to vanish at $\tan \beta = 1$. In both plots, the gray dashed line indicates $\tan \beta = 1.4$, the approximate lower bound for perturbativity of the top Yukawa coupling at the unification scale $\sim 10^{16} \GeV$.}\label{fig:BRtoh}
\end{figure}

\noindent\emph{States uncharged under the $Z$}
\\\indent At the renormalizable level, one simple scenario for obtaining large branching ratios of the NLSP to Higgses is to have both the LSP and NLSP be formed mostly from states that do not couple to the $Z$ or the photon (i.e. states with $T^3 = Y = 0$). This is achieved in the MSSM with a neutral wino-like NLSP and a bino-like LSP (as studied in ~\cite{Byakti:2012qk,Ghosh:2012mc,Baer:2012ts}). When $\mu \gg M_1, M_2$, the mixing of the NLSP and LSP with the higgsino is small. Decays of the NLSP to the Higgs are suppressed by one power of this mixing, while decays to the $Z$ require mixing of both the NLSP and LSP states and are suppressed by two powers of the mixing. This is illustrated diagramatically in fig.~\ref{fig:MixingFeyn}. From an effective operator viewpoint, integrating out the higgsinos generates the dimension five operator $(h^\dagger \sigma^3 h)\tilde{W}^0\tilde{B}$, while decays to the $Z$ will arise only from the dimension six operator $(h^\dagger\sigma^3 D_\mu  h)\tilde{W}\sigma^\mu\tilde{B}$ (since the dimension five operator does not involve couplings to the neutral goldstone mode of the Higgs) and are therefore suppressed by additional powers of the higgsino mass. The dimension five operator $\tilde W\sigma^{\mu\nu}\tilde{B}W_{\mu\nu}$ can be generated by loops involving sfermions and allows decays to the photon or $Z$, but the loop factor suppression renders this negligible. Fig.~\ref{fig:BRWtoh} shows the branching ratio of a 175 GeV wino to a Higgs and a nearly massless bino as a function of $\mu$ and $\tan \beta$, for a Higgs mass of 125 GeV in the decoupling limit. Though large $\tan \beta$ suppresses the wino decay to the Higgs, it can be highly dominant even for relatively light higgsinos.

In extensions of the MSSM there can be additional states uncharged under the $Z$ but with renormalizable couplings to the Higgs and higgsinos. This can cause the NLSP to decay dominantly to the Higgs by the same mechanism as above. Examples include the singlino $S$ of the NMSSM \cite{Das:2012rr} and string photini~\cite{Photini,PhotiniShadow}.  A singlino LSP mixes through the operator $\lambda \tilde{S}(\tilde{H_u}H_d + \tilde{H_d}H_u)$, and so unlike the bino has a decay width to the Higgs that increases with large $\tan\beta$ and is suppressed by low $\tan\beta$. String photini are superpartners of light $U(1)$ gauge fields from string theory~\cite{Photini}, which can have small mixings with the bino. A string photino LSP therefore couples to the Higgs exactly as the bino does but with a coupling suppressed by the wino-photino mixing $\epsilon$. A similar enhancement of the Higgs branching ratio can also occur for decays of a wino or bino NLSP to the goldstino of an $R$-symmetric SUSY breaking sector \cite{Thaler:2011me}.

Since the LSP can be weakly coupled to the MSSM fields in these scenarios, we will consider the limit in which all of the other MSSM fields decay promptly to the bino or neutral/charged winos, which then decay to the LSP. We restrict our attention to the case where the decay to the LSP is not displaced on detector scales ($c\tau\lesssim0.1\text{~mm}$), ensuring accurate reconstruction of collider events and avoiding constraints from searches for metastable particles.

\noindent\emph{Higgsino NLSP and gravitino LSP}
\\\indent Another scenario for obtaining large NLSP branching ratios to the Higgs can occur for a higgsino-like NLSP decaying promptly to a gravitino LSP in models with low-scale SUSY breaking. Refs.~\cite{Matchev:1999ft,ChiAtTeV,ChiAtLHC,Kats:2011qh} discuss this possibility and some of the collider signatures at the LHC and Tevatron. If $\text{sign}(\mu(\frac{M_1}{M_2}+\tan^2 \theta_W))<0$, then as $\tan \beta$ approaches unity the coupling of the lightest higgsino to the $Z$ approaches zero~\cite{ChiAtTeV}, due to cancellation of the contributions from the $\tilde H_u$ and $\tilde H_d$ components. The higgsino LSP can then decay dominantly to a Higgs and gravitino, as illustrated in fig.~\ref{fig:BRHtoh}. Since the gravitino LSP can be very weakly coupled, we will again consider the limit when all other MSSM fields decay promptly to the NLSP, which then decays promptly to the gravitino (requiring $\sqrt{\langle F \rangle} \lesssim 100\TeV$).

\noindent\emph{Spectra}
\\\indent In the models discussed above the NLSP is often accompanied by other nearly degenerate states that are important for phenomenology. We identify three classes of spectra (illustrated in fig.~\ref{fig:DirectSpectra}) that capture the possible signals:

\begin{figure}
		  \centering
        \subfigure[]{
                \includegraphics[width=1.8in]{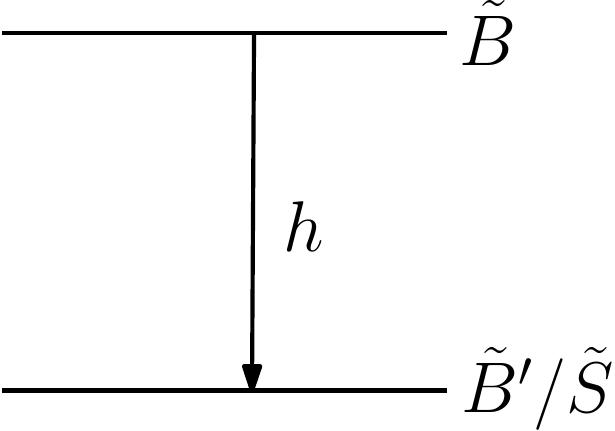}
                \label{fig:BinoNLSPDirect}
		  }
               \hspace{5mm} 
        \subfigure[]{
                \includegraphics[width=1.8in]{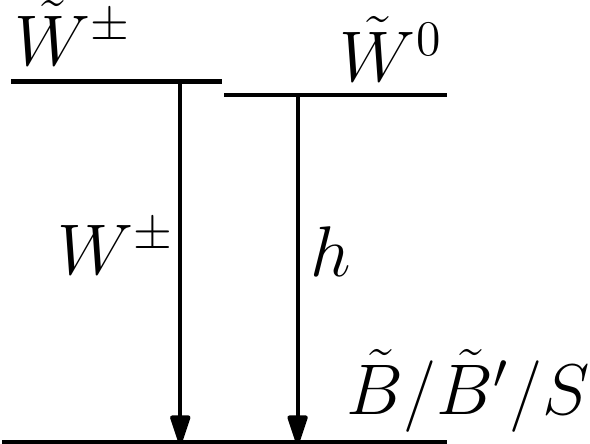}
                \label{fig:WinoNLSPDirect}
		  }
               \hspace{5mm}
        \subfigure[]{
                \includegraphics[width=1.8in]{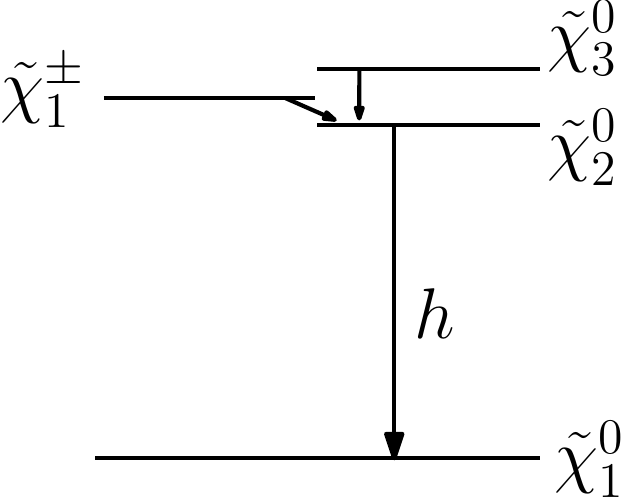}
                \label{fig:QuasiDirect}
		  }
\caption{Three types of spectra for the NLSP, LSP and associated states such that the NLSP decays dominantly the Higgs. See the text (sec.~\ref{sec:ChiToHiggs}) for discussion of each case.
}         
\label{fig:DirectSpectra}
\end{figure}

\begin{figure}
		  \centering
        \subfigure[]{
                \includegraphics[width=1.8in]{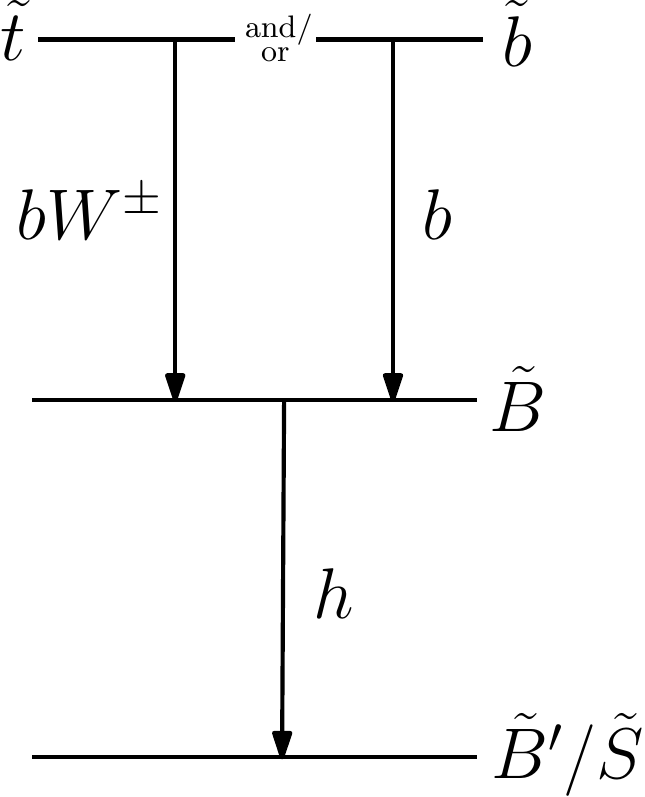}
                \label{fig:BinoNLSP}
		  }
               \hspace{5mm} 
        \subfigure[]{
                \includegraphics[width=2in]{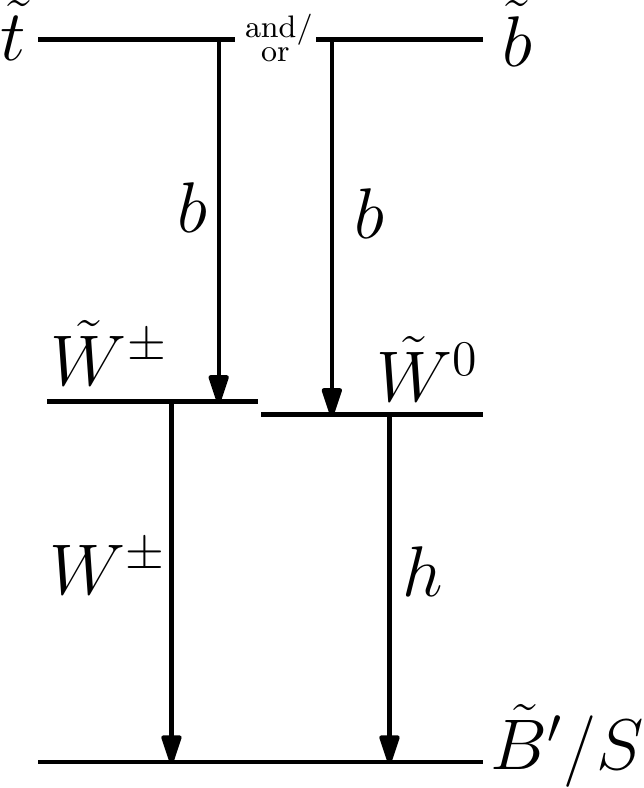}
                \label{fig:WinoNLSP}
		  }
               \hspace{5mm}
        \subfigure[]{
                \includegraphics[width=1.8in]{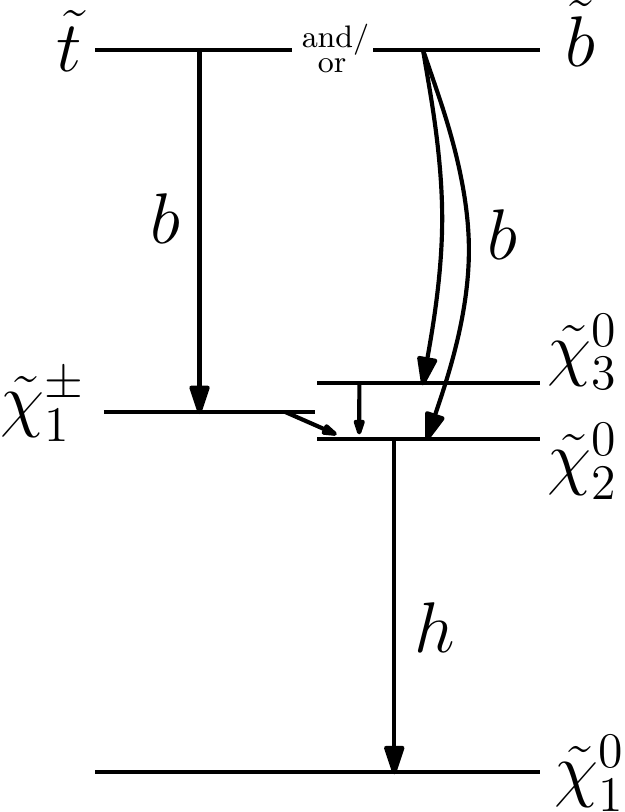}
                \label{fig:Quasi}
		  }
\caption{The decay modes of light stops and sbottoms for the three types of NLSP spectra in fig.~\ref{fig:DirectSpectra}. See the text (sec.~\ref{sec:ColoredProd}) for discussion.
}         
\label{fig:Spectra}
\end{figure}

\begin{enumerate}[(a)]

\item{\bf Bino NLSP} If the bino is the NLSP, then it can decay dominantly to a Higgs and singlino (or string photino etc.) LSP provided that $\mu$ is large. If the LSP is very weakly coupled, essentially all SUSY cascade decays will proceed through the NLSP and produce a Higgs. A pure bino has vanishing pair-production cross section, so this spectrum alone is not observable. 

\item{\bf Wino co-NLSPs} If the NLSP is a neutral wino with small mixings (large $\mu$ and $M_1$ not too close to $M_2$), then the charged winos are nearly degenerate with the NLSP. The decay of the charginos to the NLSP is then extremely suppressed; if this is the dominant decay mode then the charginos have a macroscopic decay length. We instead consider the case where the chargino decays promptly to the LSP, e.g. $\tilde W^\pm \to \tilde B W^\pm$. In this scenario not all SUSY decays will proceed through the true NLSP (the neutral wino) and produce a Higgs.   

\item{\bf Quasi-degenerate charginos} Two types of models can give the spectrum and decays shown in fig.~\ref{fig:Quasi}. One is the Higgsino NLSP scenario described in sec.~\ref{sec:ChiToHiggs}. For small mixing, the neutral Higgsino NLSP is accompanied by a chargino and a slightly heavier neutralino. We will consider the minimal case where the splitting between these states and the NLSP is $\sim 1-10 \GeV$, so that they decay promptly to NLSP but the extra objects produce in the decay are soft and do not contribute significantly to collider signals. For a Higgsino-like NLSP at $\lesssim 200$ GeV, this is achieved with $M_1, M_2 \gtrsim 500 \GeV$. Because the gravitino is very weakly coupled, the other higgsinos will always decay to the NLSP.

A second source for this spectrum are models where $M_1 \approx M_2$ so that the wino and bino are significantly mixed, while the LSP is an additional state such as a singlino or string photino. The spectrum then includes a chargino and heavier wino-bino state in addition to the mixed wino-bino NLSP. The splitting between these heavier states and the NLSP can be large enough to allow prompt decays to the latter (unlike the pure wino case).  While the wino-bino mixing is suppressed at the larger values of $\mu$ required for the Higgs decay to dominate, sufficient mixing can still be obtained for quasi-degenerate gauginos with $|M_1 - M_2|\lesssim 50$ GeV. 

\end{enumerate}

\subsection{Production through third generation squarks} \label{sec:ColoredProd}

SUSY naturalness motivates light stops $\tilde t_1$ and $\tilde t_2$, possibly with large mixing, which are generally accompanied by a light left-handed sbottom. When $m^2_{\tilde q_{3L}}$ is small and the stop and sbottom mixing are negligible, the lightest squark is $b_L$, and the tree-level splitting with $t_L$ is given by
\be
	m^2_{t_L} - m^2_{b_L} \approx m_t^2 + (1 - \sin^2\theta_W) \cos (2\beta) m_Z^2
	\label{eq:qlmass}
\ee
When $m_{b_L}\sim 300\GeV$, this gives a splitting $m_{t_L} - m_{b_L}\sim 40\GeV$. A light right-handed sbottom $\tilde b_R$ is not directly motivated by naturalness, but can result from UV realizations of natural spectra. 

The decay cascades of the lightest third generation squarks to a neutralino NLSP involve the production of a $t$ quark (on or off shell) or $b$ quark. When the NLSP decay produces a Higgs dominantly, the final states of interest will be $2h + 2(b/t)+\MET$. We find that final states with Higgses \emph{and} tops are substantially more constrained by existing leptonic SUSY searches, so we will focus on spectra where the decays of the lightest squark(s) proceed dominantly to the $2b$ final state. 

For each of the three NLSP spectra described in sec.~\ref{sec:ChiToHiggs}, we discuss the signals from the various third-generation squarks: 

\begin{enumerate}[(a)]

\item{\bf Bino NLSP} If the bino is the NLSP and the LSP is very weakly coupled, then the squarks will decay to the bino, producing a $b$ quark and, for a stop, a $W$ boson. Sbottoms therefore give $2b + 2h$ final states, while stops give $2b + 2W + 2h$. The latter case is strongly constrained by same-sign dilepton searches (see sec.~\ref{sec:ConstraintHeavierQ}), requiring the stops to be heavier than $\sim 400 \GeV$. A left-handed sbottom is thus generally required to be $\gtrsim 350 \GeV$, though a light right-handed sbottom $b_R$ can remain to give $2b + 2h$ at observable levels.

\item{\bf Wino co-NLSPs} If the winos are co-NLSPs, the squarks can decay to either the charged or neutral wino. This decay will be dominant if the LSP is a very weakly coupled state such as a string photino. If the splitting between the winos and the squarks is not too much larger than the top mass, then decays to bottom quarks dominate, so that sbottoms decay to the neutral wino and stops to the charged wino. Sbottom decays through the winos then give the final state $2b + 2h$, while stops give $2b + 2W$. The latter is constrained by direct stop searches (sec.~\ref{sec:ConstraintHeavierQ}), requiring the stops to be heavier than $\sim 350 \GeV$. 

If the LSP is a bino, then the squarks can decay directly to the LSP and a quark. This will be the dominant decay for right-handed squarks, so Higgses will not be produced in this case. Left-handed squarks will have a subdominant branching ratio to the bino (at least $\sim 5\%$), but this is enough for searches in $3b + \MET$ final states to be sensitive, as will be shown in sec.~\ref{sec:ConstraintBRtoLSP}. These searches constrain the left-handed sbottom to be heavier than $\sim 350 \GeV$. The bino LSP scenario therefore does not have as much potential to give observable signals during the 8 TeV run.

\item{\bf Quasi-degenerate charginos} As before, we will assume that the splitting between the NLSP and the squarks is small enough that the squarks decay dominantly to bottoms rather than tops. The sbottoms then decay to neutralinos while stops decay to charginos. All of these decays eventually produce the NLSP, possibly with additional soft decay products, so all squarks give the $2b + 2h$ final state. 

\end{enumerate}

\subsection{A Simplified Model}\label{sec:SimpModel}
\begin{figure}
\includegraphics{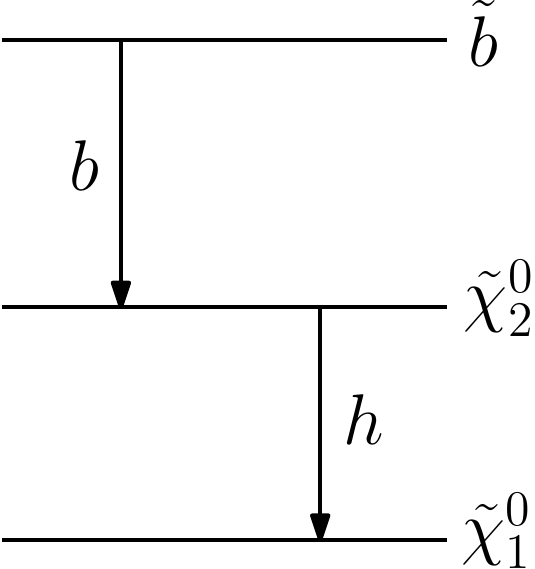}
\caption{The simplified model we consider to capture a wide range of possibilities for third generation squarks producing Higgses in cascade decays. Only pair production of the bottom squark $b$ is considered.
}         
\label{fig:SimpModel}
\end{figure}

From the above discussion, we can see that the signatures of a third generation squark decaying to the $2b+2h$ final state can in many cases be well described by the simplified model (illustrated in fig.~\ref{fig:SimpModel}) of a sbottom decaying with 100\% branching fraction to a $b$ quark and neutral NLSP which decays with 100\% branching fraction to the Higgs and LSP, giving the final state $2h + 2b + \MET$. Some of the above scenarios give additional final state particles that carry very little energy due to small mass splittings. These will generally not appear as hard objects in collider events and will only slightly affect the kinematics of the hard $b$'s and Higgses. Therefore the simplified model is an appropriate description of the collider signatures in these models as well. 

The pair production cross sections are essentially the same for all sbottoms and stops (up to small electroweak effects), so this model is a good description for either a light stop or sbottom. The model is described by two parameters, the sbottom/stop mass and the NLSP mass. For simplicity, we fix the LSP mass at $1\GeV$  and consider a 125 GeV lightest Higgs boson in the decoupling limit, where its couplings are equal to those of the SM Higgs. 

We also note that several other types of models can give enhanced Higgs production with similar final states, i.e. Higgses + ($b$-quarks) + (low MET). For example, in models with additional vector-like generations, decays of the vector-like quarks to a Higgs and a top or bottom also produce this final state. In particular, decays of a new bottom-like quark to $bh$ give the same event kinematics as this simplified model with NLSP mass equal to the Higgs mass, though with different production cross section. The phenomenology of top-like quarks is explored in~\cite{Azatov:2012rj}. Ref.~\cite{Brust:2012uf} showed that for $R$-parity violating SUSY, Higgs bosons can be produced preferentially in the decays between third generation squarks, e.g. $\tilde t_2 \rightarrow \tilde t_1 h$, giving the same signatures. Another possibility is new physics enhancing the rare decay $t\to ch$, which can be observed in top pair production events \cite{Craig}. Other recent work studied the possibility that light colored particles in loops could increase the rate of Higgs pair production $pp \to hh$ to make its effects comparable to the SM single Higgs production rate~\cite{Kribs:2012kz}, which would have similar signatures in multilepton and diphoton searches. We expect that some of our conclusions based on this simplified model will also apply qualitatively to these more general models, as discussed in sec.~\ref{sec:Conclusions}. 

\begin{figure}
		  \centering
        \subfigure[]{
                \includegraphics[width=3in]{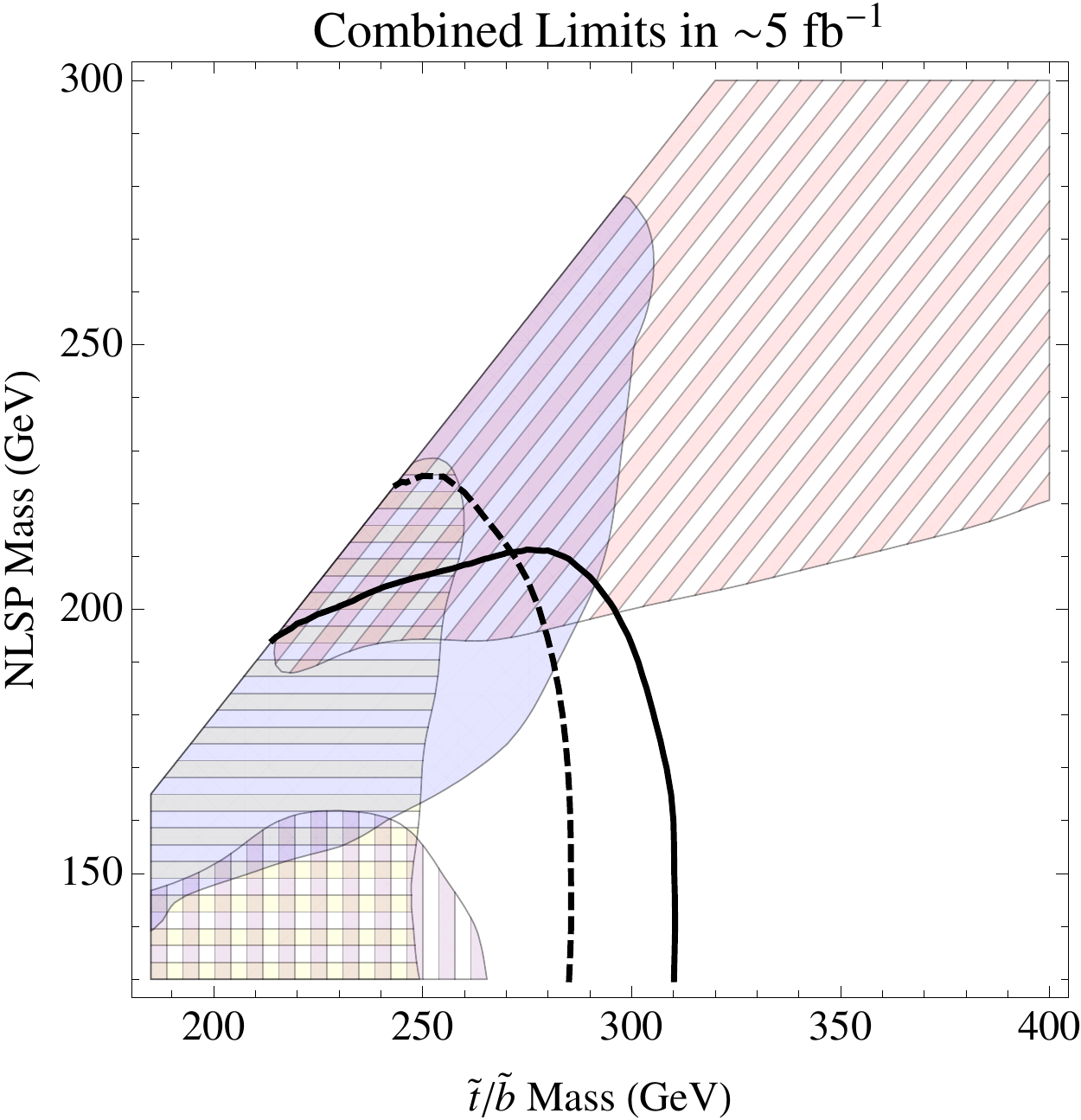}
                \label{fig:CombinedLimits5}
		  }
               \hspace{5mm} 
 	\subfigure[]{
                \includegraphics[width=3in]{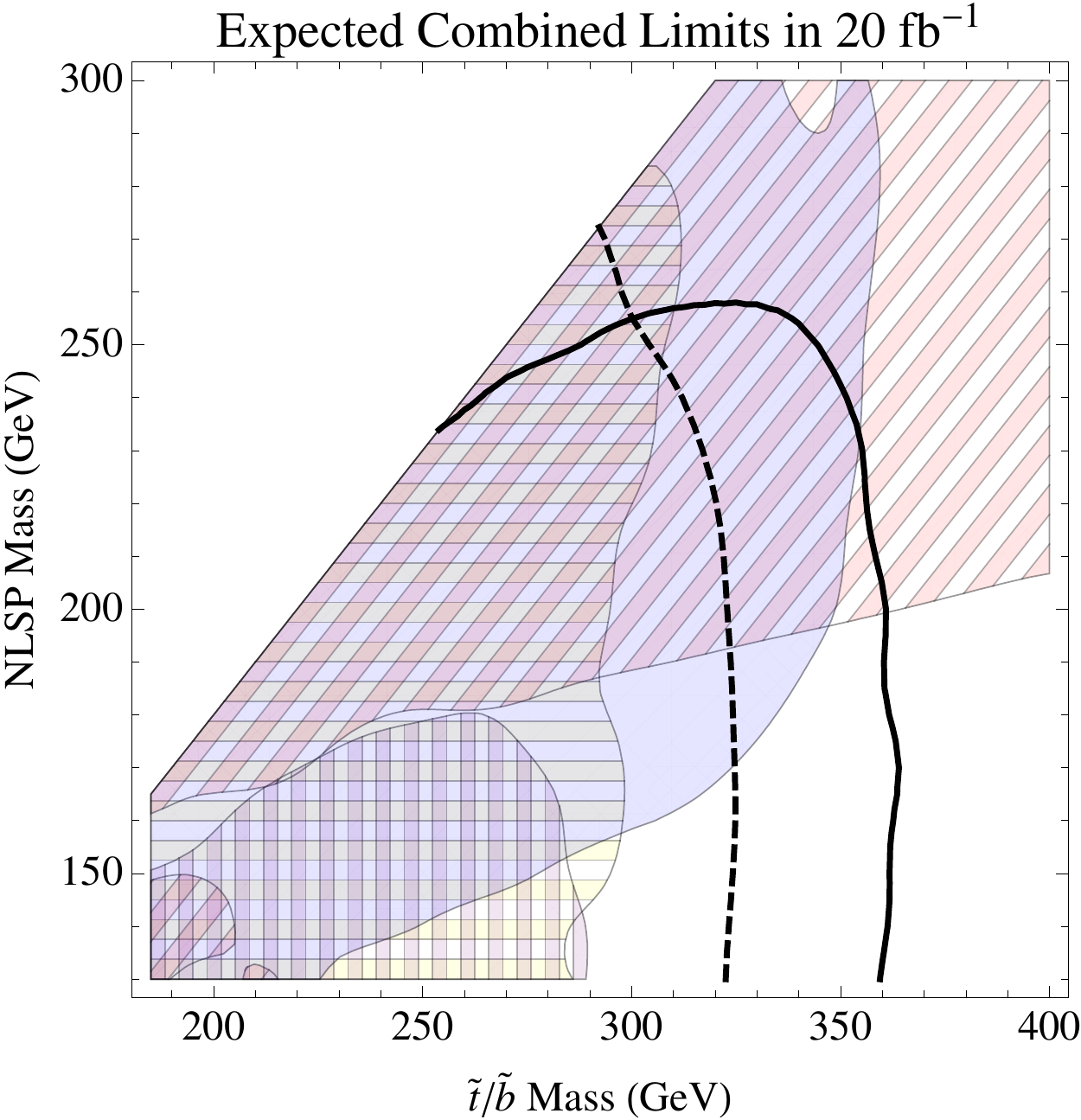}
                \label{fig:CombinedLimits20}
		  }
               \hspace{5mm} 
        \caption{The 95\% excluded regions for the simplified model of sec. \ref{sec:SimpModel} from (a) the existing searches in $\sim5\fb^{-1}$ discussed in sec. \ref{sec:Constraints} and \ref{sec:GG}, and (b) the \emph{expected} 95\% exclusions in $20\fb^{-1}$ at 8 TeV for the same searches, extrapolated as described in the text. The diagonal hatching indicates the region constrained by the CMS8 $3b+\MET$ ($\alpha_T$) search \cite{3bCMS}, horizontal for the CMS7 multilepton search \cite{Chatrchyan:2012ye},  vertical for the CMS8 same-sign dilepton + 2$b$ search \cite{SSbjet8}, and solid for the CMS8 $\gamma\gamma+\MET$ search \cite{FermiophobicCMS8}. The black solid (dashed) contours are the expected 95\% exclusions in $5\fb^{-1}$ and $20\fb^{-1}$ from a proposed search in $\gamma\gamma$ in association with at least two (one) $b$-tagged jets, described in sec. \ref{sec:ggb}. \label{fig:CombinedLimits}}
\end{figure}

In the following sections, we describe in detail the constraints and signals of this simplified model in existing SUSY searches (sec. \ref{sec:Constraints}) and existing Higgs diphoton searches (sec. \ref{sec:Diphoton}). These results are summarized in fig. \ref{fig:CombinedLimits}. In short, we find that a sbottom heavier than $\sim 260\GeV$ with an NLSP lighter than $\sim200\GeV$ is not currently excluded. In this region there can be strong signals in a proposed search for resonant $\gamma\gamma+1/2b$ (sec. \ref{sec:ggb}), and even observable enhancements in the inclusive Higgs diphoton searches (secs. \ref{sec:GG} and \ref{sec:OtherSearch}). We also consider the constraints on deviations from the simplified model, in particular non-zero branching ratios of the NLSP to the $Z$ (sec. \ref{sec:ConstraintBRtoZ}), non-zero branching ratio for the direct decay of the squark to the LSP (sec. \ref{sec:ConstraintBRtoLSP}), and decays of the heavier third generation squarks (sec. \ref{sec:ConstraintHeavierQ}); we find that the simplified model is a good description of the dominant LHC signals for a wide range of realistic models.

\section{Constraints from SUSY searches} \label{sec:Constraints}

In this section we determine the constraints on models like those discussed above from existing searches for supersymmetry. To do so (and to determine signal rates in Higgs searches in sec.~\ref{sec:Diphoton}), we perform Monte Carlo (MC) simulation of collider events to determine the efficiencies of various LHC searches for these models. We use {\tt MadGraph~5}~\cite{Alwall:2011uj} to generate SUSY production events, {\tt Pythia~6.4}~\cite{Sjostrand:2006za} to decay particles, perform parton showering and hadronization and produce initial and final state radiation, and {\tt PGS4} \cite{PGS4} to simulate object reconstruction in detectors (we use the {\tt Pythia-PGS} package to interface these tools). We use the default {\tt PGS} parameter cards for the ATLAS and CMS detectors, but to implement $b$-tagging efficiencies we use truth-level $b$-jets and apply the $p_T$-dependent $b$-tagging efficiencies reported by the ATLAS and CMS experiments for each search. We scan a grid of points in $M_{\NLSP}$ and $M_{\tilde t/\tilde b}$ spaced by 5 to 15 GeV, and generate 50,000 events at each point for the diphoton channels and 250,000 events for the non-diphoton channels to obtain adequate statistics for low rate searches. We obtain the cross sections for production of third-generation squarks from \cite{SquarkProd} and for neutralino/chargino direct production from {\tt Prospino 2.1}~\cite{Beenakker:1999xh}; in both cases we use the central value of the prediction and do not account for the theoretical uncertainty (giving more conservative bounds). We take SM Higgs boson cross sections and branching ratios from~\cite{LHCHiggsCrossSectionWorkingGroup:2011ti}.  When available we use the $95\% \CLs$ visible cross section limits presented in the searches we study; when these are not available we compute the $95\% \CLs$ limit from the observed event rate and backgrounds as described in~\cite{Junk:1999kv}. Where not otherwise noted, we estimate the expected limits in $20\fb^{-1}$ by linearly extrapolating the expected background rate and systematic uncertainty from the current results.

Where possible, we have validated our results against quoted limits and generally find agreement within $\sim 25\%$. Nonetheless there is the possibility of significant systematic errors since the detector simulation we use does not fully capture the performance of the ATLAS and CMS detector, in particular in analyses that select for large missing energy and are therefore sensitive to the tails of distributions. For these reasons we emphasize that the results of this section should not be taken as strict experimental bounds but rather as approximate guidelines as to what regions of parameter space are accessible by various searches.

We first consider constraints on the simplified model of section~\ref{sec:SimpModel}. We then discuss the realm of validity of the simplified model, considering further constraints on models with nonzero branching ratio of the NLSP to the $Z$ and of the lightest squark directly to the LSP, as well as constraints on the decays of the heavier 3rd generation squarks. 

\subsection{Constraints on the simplified model from SUSY searches} 
\label{sec:ConstraintSimpModel}

\begin{figure}
\includegraphics[width=3in]{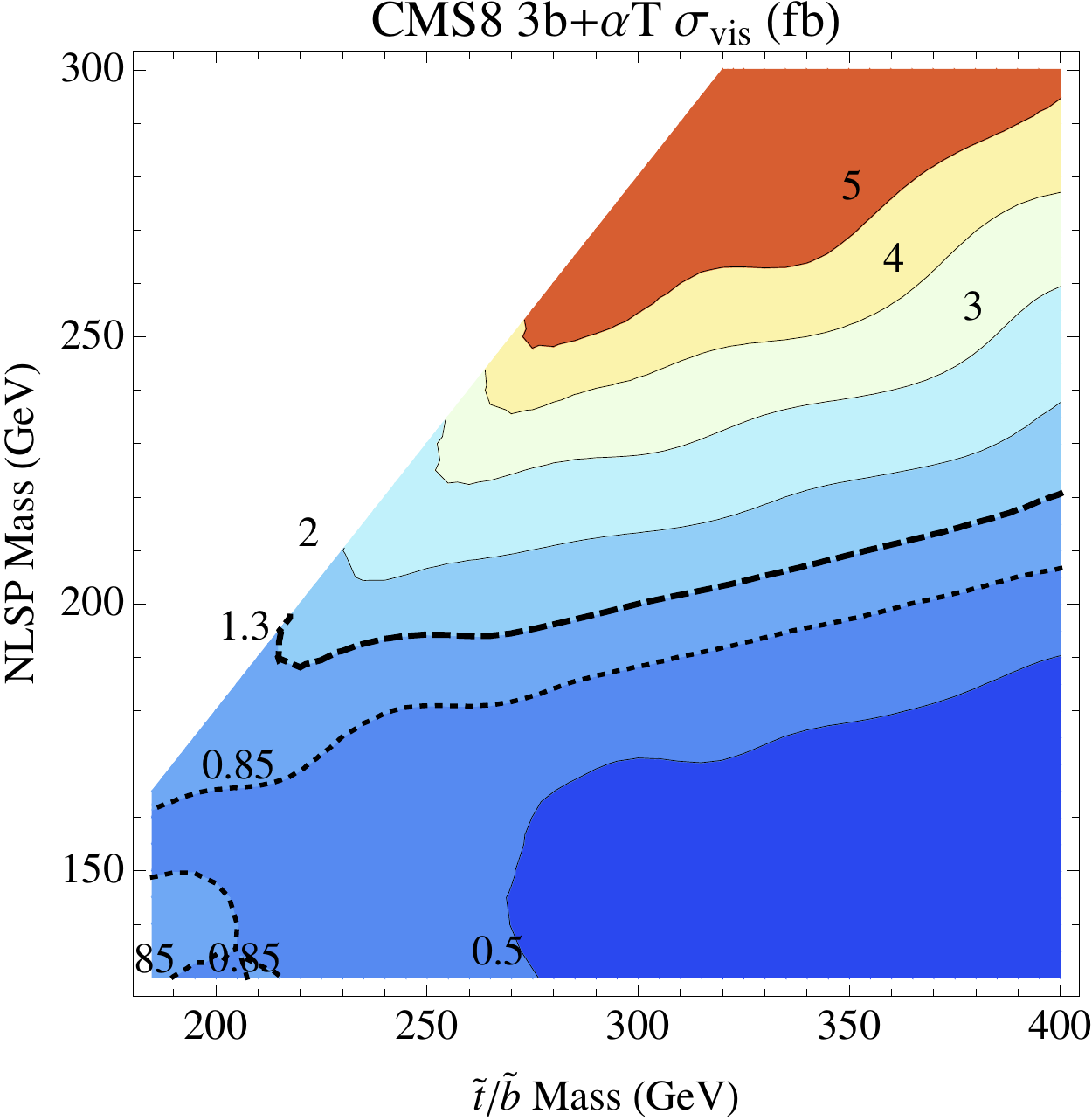}
\caption{{\bf CMS $\mathbf{3b+\MET}$ ($\mathbf{\alpha_T}$):} Visible cross section from our simplified model for the CMS search in final states with 3 $b$-jets and missing energy~\cite{3bCMS}. The thick dashed contour is the observed 95\% limit, and dotted is the expected limit extrapolated to $20\fb^{-1}$. 
}         
\label{fig:3bCMSConstraint}
\end{figure}

\noindent\emph{CMS $3b+\MET$ $(\alpha_T)$}\\
\indent The final states from our simplified model contain at least two $b$-quarks, a moderate amount of missing energy, and the decay products of the two Higgses. A 125 GeV Higgs has a branching ratio to bottom quarks of 58\% in the SM, so events can easily contain up to six $b$-jets. Standard Model processes rarely produce more than two bottom quarks in an event, so searches in final states with three or more $b$-jets can place strong limits on new physics. Both the ATLAS and CMS collaborations have performed searches in final states with three or more $b$-tagged jets and missing energy ~\cite{3bATLAS,3bCMS}. The CMS search in $3.9 \fb^{-1}$ of 8 TeV data places the strongest individual limit; in the region $H_T>325 \GeV$, $11.7$ events were expected and $6$ were observed, for which we compute a 95\% $\text{CL}_S$ limit on the visible cross section of $1.3 \fb$. The results from our MC simulation for the visible cross section (total BSM cross section times branching ratio to final state times selection/identification efficiency) in this CMS search channel is shown in fig.~\ref{fig:3bCMSConstraint}.  The bound is satisfied for NLSP masses $\lesssim 200 \GeV$. The ATLAS $3b$ search places weaker bounds and is not shown.

\noindent\emph{CMS SSDL + $2b$}\\
\indent Another type of analysis that can place relevant constraints are searches in final states with $b$-jets and same-sign leptons, such as those performed by CMS in data at both 7 and 8 TeV~\cite{SSbjet7,SSbjet8}. 

\begin{figure}
\includegraphics[width=3in]{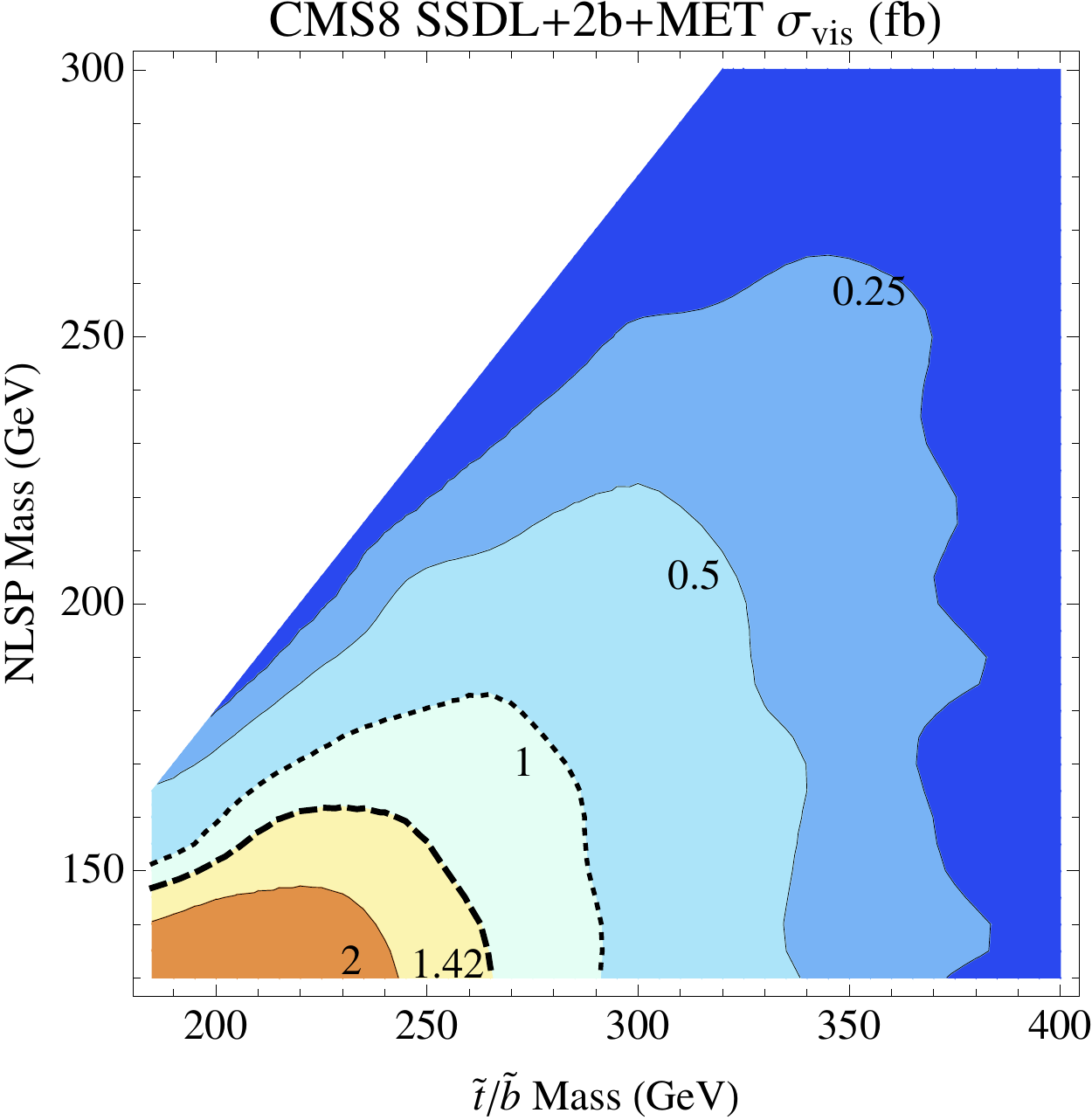}
\caption{{\bf CMS SSDL + $\mathbf{2b}$: }Visible cross section from our simplified model for the cuts of signal region 2 of the CMS search in final states with 2 $b$-jets and same-sign leptons at 8 TeV~\cite{SSbjet8}. The thick dashed contour is the observed 95\% limit, and dotted is the expected limit extrapolated to $20\fb^{-1}$.
}         
\label{fig:SSl2bCMS8Constraint}
\end{figure}

This too is a rare final state in the Standard Model. In our models, same-sign leptons can be produced if both Higgses in the event produce a lepton in their decay. This rate is suppressed by small branching fractions, but unlike the 3$b$-jet searches this analysis does not require large MET or $H_T$. The strongest constraint on our model comes from signal region 2 of the 8 TeV search~\cite{SSbjet8}, which required two $b$-tagged jets and two positively charged leptons as well as MET $> 30 \GeV$ and $H_T > 80 \GeV$. Through an apparent downward fluctuation, zero events were observed in this channel though 11 were observed in the corresponding channel with two negatively charged leptons. A 95\% CL upper limit of 2.8 expected events from non-SM processes was placed using the $\CLs$ method, corresponding to a visible cross section of .71 fb at 8 TeV, which we interpret as a limit of 1.42 fb in the combined $++/--$ channel (which would be populated evenly in our simplified model). The predicted cross sections from our simplified model are shown in fig.~\ref{fig:SSl2bCMS8Constraint}.  For an NLSP mass $\lesssim 160 \GeV$, the squark mass is bounded to be $\gtrsim 260 \GeV$.

We validated our implementation of this search by computing efficiencies for the simplified model of $p p \to \tilde{b}\tilde{b}, \tilde{b}\to t \chi_1^{\pm}, \chi_1^{\pm}\to \chi_1^0 W^{\pm}$;  the efficiencies we calculated corresponded to limits $20\%$ stronger than reported in ref.~\cite{SSbjet8}.

\begin{figure}
\includegraphics[width=3in]{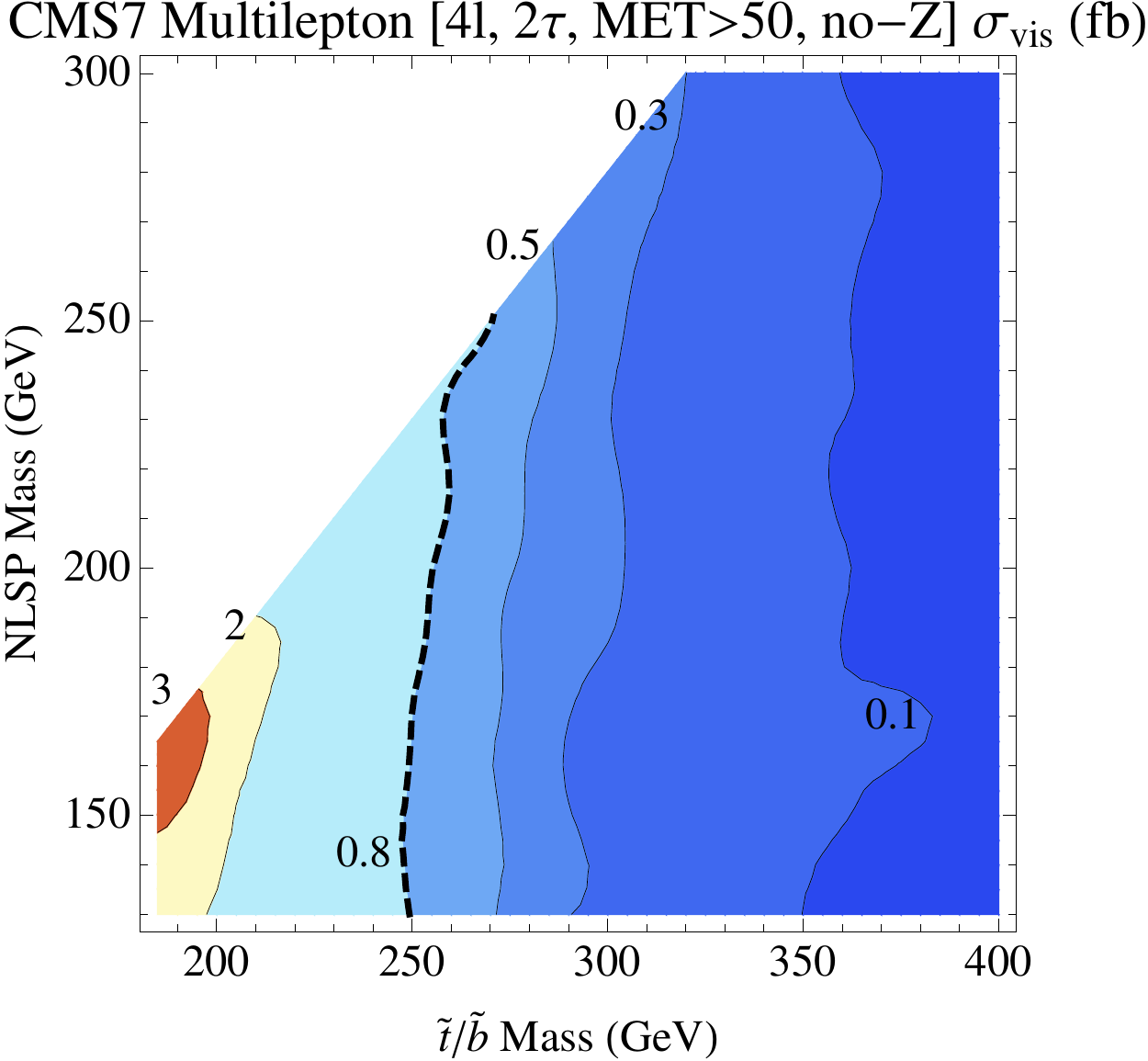}
\caption{{\bf CMS Multilepton:} Visible cross section from our simplified model for the combined $4l$, 2$\tau_h$, MET $> 50 \GeV$, no-$Z$ channels in the CMS 7 TeV multilepton search~\cite{Chatrchyan:2012ye}. The thick dashed contour is the observed 95\% limit.
}         
\label{fig:multileptonCMS7Constraint}
\end{figure}

\noindent\emph{CMS Multilepton}\\
\indent Multilepton searches without $b$-tags also put a strong constraint on our model, especially when identifying tau pairs, which are produced with branching fraction 6.4\% in the decay of a 125 GeV Higgs.
CMS has a performed search in three ($3\ell$) and four lepton ($4\ell$) channels in $4.98 \fb^{-1}$ of 7 TeV data~\cite{Chatrchyan:2012ye}. Although our models have slightly higher rates in the $3\ell$ channels, much lower backgrounds make the $4\ell$ channels far more constraining. The most sensitive $4\ell$ channels are those vetoing $Z$ resonances (`no-$Z$'), which are populated by our model when both Higgses decay either to $WW^*$ or $\tau^+\tau^-$. The search channels are further separated by the number of hadronic taus ($0/1/2\tau$) reconstructed. The branching ratios to the $1/2\tau$ channels are several times larger than the $0\tau$ channel while the background remains low. The dominant constraint over most of our parameter space comes from the $4l$, 2$\tau$, MET $>$ 50 GeV, no-$Z$ channels. Combining the low and high $H_T$ channels, in $4.98 \fb^{-1}$ $2.2\pm 0.9$ events were expected and 1 was observed, from which we compute a $95\%$ $\text{CL}_S$ limit on the visible cross section of 0.8$\fb$. Fig.~\ref{fig:multileptonCMS7Constraint} shows the visible cross section for our simplified model. The rate depends only weakly on the NLSP mass and the limit is satisfied for squark masses $\gtrsim 250 \GeV$. We note that the $0\tau$ and $1\tau$ regions with the same kinematic selections would also set a strong limit but for an observation of 4 events where $\sim1$ was expected. Taken together with the $2\tau$ channels, the observed rate would be consistent with a $\sim250$ GeV squark in our simplified model. We estimate the expected limit in $20\fb^{-1}$ of 8 TeV data (shown in fig. \ref{fig:CombinedLimits}) by applying the same search cuts to the 8 TeV signal while assuming the same background rate as in the 7 TeV search, scaled only by the luminosity.

Refs. \cite{ContrerasCampana:2011aa} consider the limits from the same multilepton search on SM $hW$ production, and ref. \cite{Craig} considers the limits set on $hW$ production in BSM physics with a focus on the rare decay $t\to ch$. Compared to $hW$  production, the multilepton limits on $hh$ production are stronger due to the increased rates in $4\ell$ channels and the importance of hadronic tau channels (not considered in refs. \cite{ContrerasCampana:2011aa,Craig}). For comparison, our simulated efficiencies for the SM Higgs contribution to the $0\tau$ channels were about 20\% higher than reported in ref.~\cite{ContrerasCampana:2011aa}, likely due to a more detailed modeling of the CMS lepton efficiencies in their study.  

Limits in the $4\ell$ channels are statistics dominated and will likely improve significantly with the full $8$ TeV run, especially with a properly combined analysis of the sensitive channels. If limits continue to improve, SM Higgs production will in fact be an important background in the most sensitive channels. The $3\ell$ channels may also become sensitive if backgrounds can be reduced by $b$-tagging, a strategy also suggested in~\cite{ContrerasCampana:2011aa,Craig} for identifying Higgs production in the $tth$ and $t\to ch$ channels.

\subsection{Constraints on NLSP branching ratio to $Z$}
\label{sec:ConstraintBRtoZ}

An ATLAS search in 2.05 $\fb^{-1}$ for $b$-jets and a leptonically decaying $Z$~\cite{Zbb} places a constraint on the branching ratio of the NLSP to a $Z$ in our model. This search targets pair production of stops in GMSB, with the stop decaying through the neutralino to a Z and gravitino, $t\to b(\chi_1^+ \to \chi_1^0) \to bZ\tilde{G}$. This is identical to our simplified model but with 100\% branching fraction to a $Z$ instead of to the Higgs. They quote a limit of $m_{\tilde t} > 310 (330)$ GeV for an NLSP mass $< (>)$ 190 GeV.

\begin{figure}
\includegraphics[width=3in]{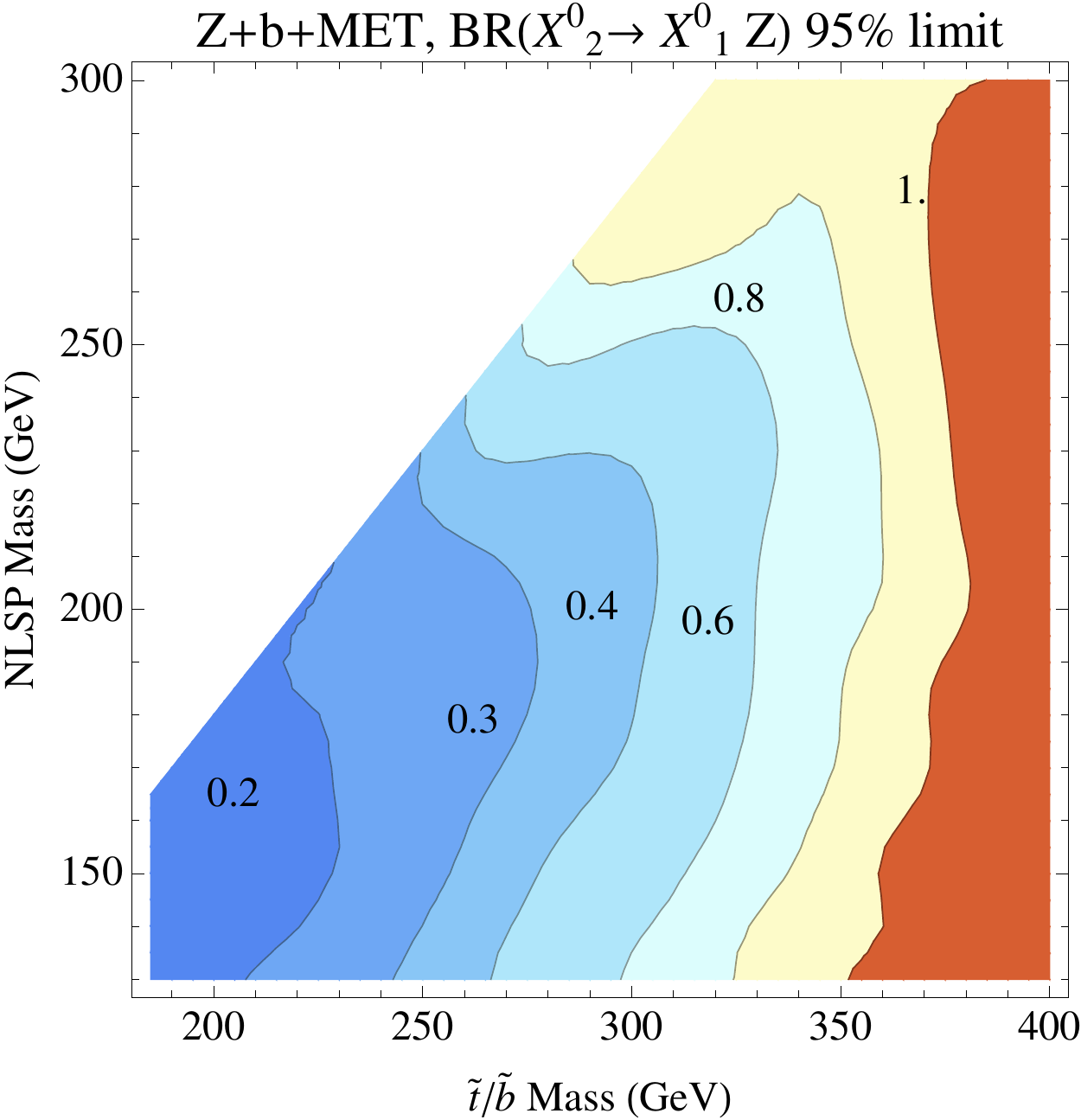}
\caption{95\% CL upper limit on the branching ratio of the NLSP to the $Z$ in the simplified model from the ATLAS search in final states with $b$-jets and a leptonically decaying $Z$~\cite{Zbb}. }         
\label{fig:NLSPtoZConstraint}
\end{figure}

We determine efficiencies for this search by generating events in which one or both NLSPs decays to the $Z$ instead of the Higgs. The visible cross section limits can be translated into a 95\% constraint on the branching ratio BR($\chi_2^0 \to \chi_1^0 Z$), as shown in fig.~\ref{fig:NLSPtoZConstraint}. Branching ratios less than these values are easily achieved in the models described in sec.~\ref{sec:ChiToHiggs}. Nominally the contours of $100\%$ should agree with the quoted experimental exclusions for the GMSB stop model, and we see that our simulation is somewhat overestimating the strength of this search.

\subsection{Constraints on squark branching ratio directly to LSP} \label{sec:ConstraintBRtoLSP} 

\begin{figure}
\includegraphics[width=3in]{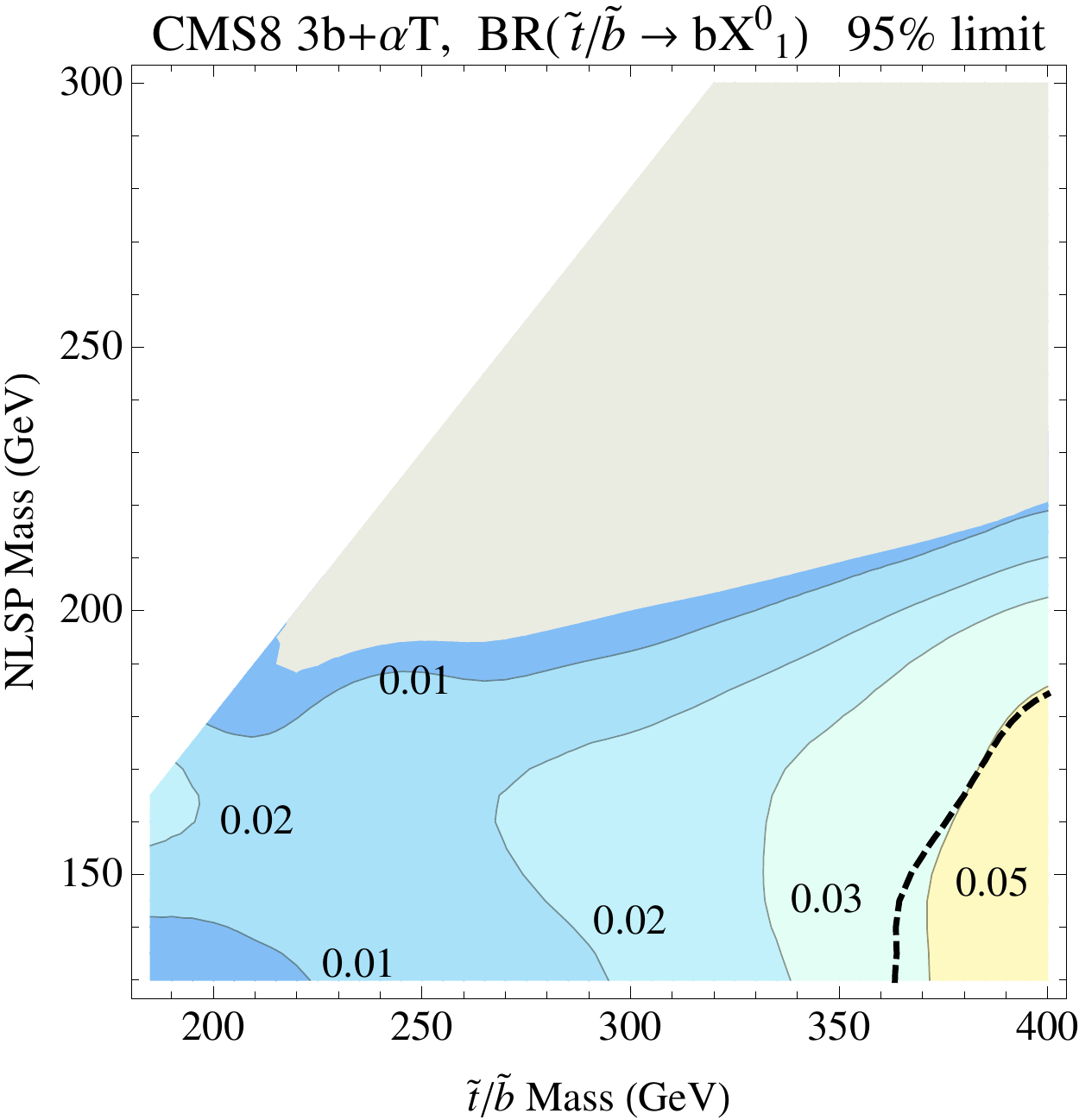}
\caption{95\% CL upper limit on the branching ratio of the squark directly to the LSP from the CMS search in final states with 3 $b$-jets and missing energy in $3.9 \fb^{-1}$ at 8 TeV~\cite{3bCMS}. The gray region is ruled out by this search even for 100\% branching fraction to the NLSP (see fig. \ref{fig:3bCMSConstraint}). The dashed contour is the 95\% exclusion for the specific model of a left handed sbottom with a wino NLSP and bino LSP.}         
\label{fig:ConstraintBRtoLSP}
\end{figure}

If the squark can decay directly to the LSP instead of through the NLSP, then the bounds from searches with MET cuts and $b$-tags become considerably tighter. Figure ~\ref{fig:ConstraintBRtoLSP} show the 95\% upper limit on the branching ratio to this mode from the CMS $3b+\alpha_T$ search, obtained by simulating events where one or both squarks decay directly to the LSP instead of through the NLSP. 

The limit on this branching ratio is 1-5\% in the region of interest. For a weakly coupled LSP (photino/singlino/goldstino/gravitino) this constraint is easily satisfied. However, for a left handed sbottom with a wino NLSP and bino LSP, as described in sec. \ref{sec:ColoredProd},  $m_{\tilde b_L}\lesssim 350 \GeV$ is ruled out.

\subsection{Constraints on heavier third generation squark decays}\label{sec:ConstraintHeavierQ}

While we have focused on the decays of the lightest squark, there are also constraints on the heavier squarks. When the lightest squark is $\tilde b_L$, the constraints on $\tilde t_L$ directly constrain the parameter space of allowed sbottom masses (see eq. \ref{eq:qlmass}), while in the other cases the limits simply set a lower bound on the masses of other squarks in the spectra.

The strongest constraints on the heavier squarks in the spectra occur when their decays give the final state $t\overline{t}hh+\MET$, which can arise for example from the decay of a heavier stop to a bino NLSP, $\tilde t \to \tilde B t$ with the $\tilde{B}$ decaying to a Higgs. In particular the CMS SSDL + 2$b$ search, discussed in sec. \ref{sec:ConstraintSimpModel}, places a much stronger limit on this final state than on the $b\overline{b}hh+\MET$ state due to leptonic decays of the $W$'s produced from the top decay. We find that a squark decaying with 100\% branching fraction to this mode with an on-shell top is excluded at 95\% for masses $\lesssim 400\GeV$ by the search in $3.9\fb^{-1}$ at 8 TeV; in $20\fb^{-1}$ the expected reach is $\sim450\GeV$.

\begin{figure}
\includegraphics[width=3in]{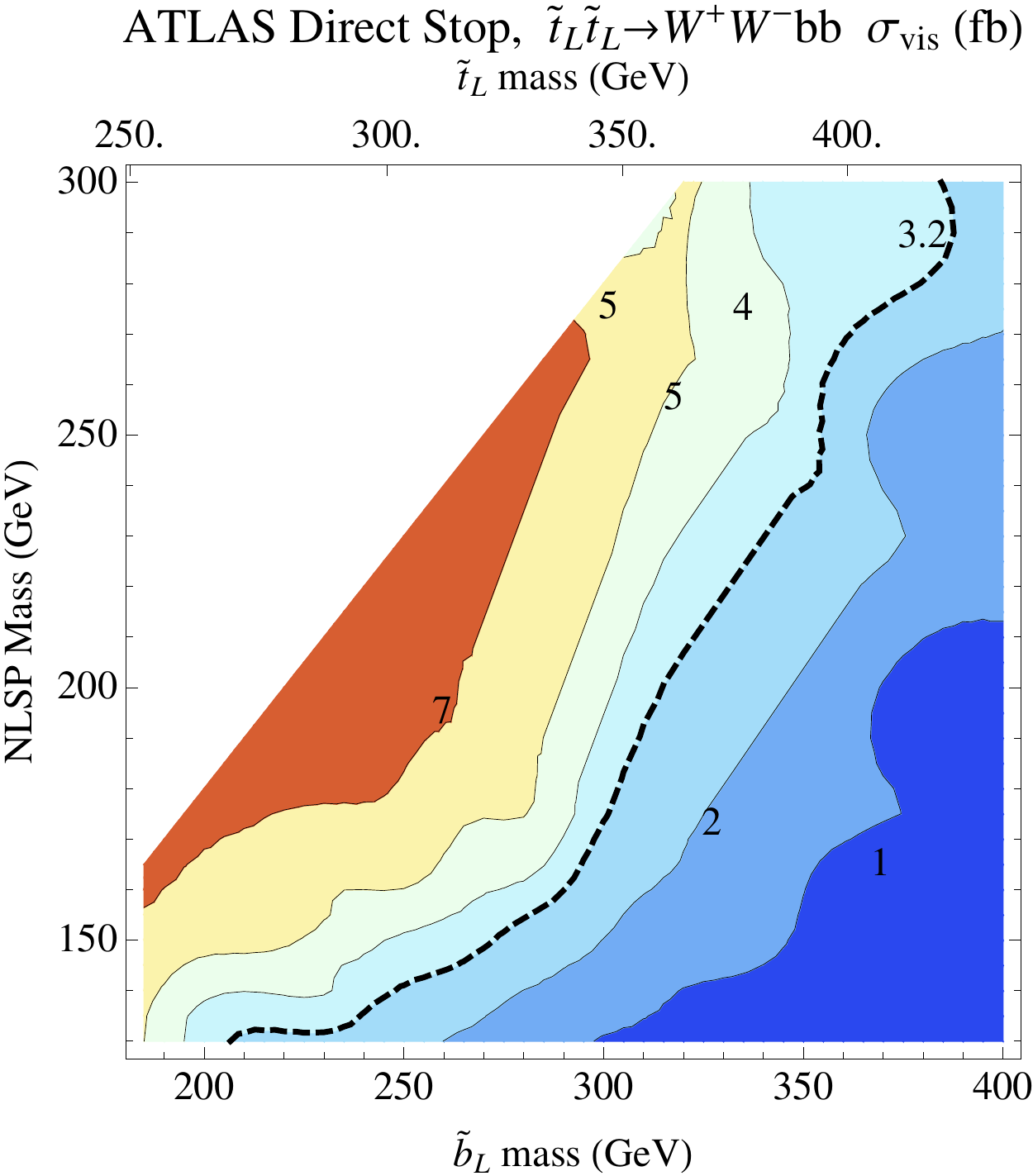}
\caption{Visible cross section for the decay $\tilde t_L \tilde t_L \to W^+W^-b\overline{b}$ in the ATLAS single lepton direct stop search \cite{ATLASstop}. $m_{\tilde t_L}$ and $m_{\tilde b_L}$ are related by eq. \ref{eq:qlmass}, for which we chose $\tan\beta=3$. The thick dashed contour is the 95\% limit from $4.9{\fb^{-1}}$ at 7 TeV.\label{fig:directstopLimits}}
\end{figure}

Direct stop searches have very targeted kinematic requirements and do not significantly constrain the $(t\overline{t}/b\overline{b})hh+\MET$ final states. However, in the case of $\tilde t_L$ decaying through a chargino co-NLSP to the final state $W^+W^-b\overline{b}+\MET$ (as in figure~\ref{fig:WinoNLSP}), the final state is similar enough to the $t\overline{t}+\MET$ topology to be constrained. The strongest limit comes from the lowest MET signal region of the ATLAS 7 TeV search for stop pair production with a single leptonic $W$ decay in $4.9{\fb^{-1}}$~\cite{ATLASstop}.  When validating our method for the stop decay model in ref. \cite{ATLASstop}, we found the ratio of our efficiencies to those presented to be $\sim0.5$ uniformly over the parameter space presented. We therefore scaled our efficiencies by a factor of $2$ to compensate. For stops near $350 \GeV$, the cross-section changes by a factor of two as the mass is changed by $\sim 40 \GeV$, so we expect our bounds on squark mass to be within this margin of error. The constraints we obtain are shown in fig.~\ref{fig:directstopLimits} in terms of both the $t_L$ mass (top of frame) and the corresponding $b_L$ mass (bottom of frame) for comparison with the direct constraints on $b_L$ production. The limit is roughly $m_{\tilde t_L}\gtrsim 350\GeV$ for an NLSP heavier than 150 GeV. For heavier $\tilde t_L$ the $t\overline{t}hh+\MET$ final state also opens up, but we find that the dominant constraint is still from the direct stop search. 

\section{Diphoton channels}\label{sec:Diphoton}

We now discuss the possibility of observing the simplified model of sec.~\ref{sec:SimpModel} in searches taking advantage of the Higgs diphoton resonance. We first describe our methods in the simplest case by considering the inclusive diphoton searches. We then discuss the potential reach of the existing CMS searches  targeting SM associated production in $\gamma\gamma+\ell$ and $\gamma\gamma+\MET$ channels~\cite{FermiophobicCMS8} and explore possible ways to increase their sensitivity to Higgs production in SUSY processes. Finally, we propose a search in $\gamma\gamma + 1/2b$ final states that could be much more sensitive to our model than other searches.

\subsection{Inclusive diphoton rates} \label{sec:GG}

Both ATLAS and CMS have presented combined searches at 7 and 8 TeV for inclusive Higgs to diphoton final states, finding respectively best fit values of $\sigma/\sigma_{\SM} = 1.9\pm0.5$ and $\sigma/\sigma_{\SM} = 1.56\pm0.43$ for a Higgs mass near $125$ GeV~\cite{DiphotonATLAS, DiphotonCMS}. 

An important point is that this search does not veto on any extra activity in the event, so we expect our model to give a fractional enhancement to the rate of roughly $\sim 2\times\sigma_{\tilde{b}\tilde{b}} / \sigma_{h,\SM}$, where $\sigma_{\tilde{b}\tilde{b}}$ is the squark production cross section and the factor of 2 arises because two Higgses are produced in each SUSY event. To obtain a more accurate result reflecting the changes in kinematics and isolation between the SM and SUSY cascade production modes, we again generated simulated signals as described in sec.~\ref{sec:Constraints} and applied the kinematic cuts specified by each search. 

For searches in diphoton final states, the total efficiencies $\epsilon$ depend on the efficiencies of the photon identification used by ATLAS and CMS. The {\tt PGS4} detector simulation applies only isolation cuts and a cut on the ratio of energy in the hadronic and EM calorimeters when identifying photons, while ATLAS and CMS use additional cuts on the shape and quality of the EM shower in the calorimeter. Our MC simulation therefore overestimates the photon identification efficiency. To account for this, we simulated $pp \to Vh \to V\gamma \gamma$ in the SM and computed the efficiency for the ATLAS~\cite{DiphotonATLAS} and CMS~\cite{DiphotonCMS} diphoton searches assuming the photon identification reported by {\tt PGS4}. We define variables $\kappa$ as the ratios of these quantities ($\epsilon_{h \to \gamma \gamma}^{\text{MC}}$) to the actual $pp \to Vh \to \gamma \gamma$ efficiencies reported by ATLAS and CMS including the tighter photon identification criteria ($\epsilon_{h \to \gamma \gamma}^{\text{ATLAS}}$, $\epsilon_{h \to \gamma \gamma}^{\text{CMS}}$):
\be 
\kappa_{\text{ATLAS/CMS}} \equiv \frac{\epsilon_{h \to \gamma \gamma}^{\text{ATLAS/CMS}}}{\epsilon_{h \to \gamma \gamma}^{\text{MC}}} 
\ee
At both 7 and 8 TeV we find $\kappa_{\text{ATLAS}}\sim\kappa_{\text{CMS}}\sim0.7$. We interpret $\kappa$ as a constant efficiency for both photons from a Higgs decay to pass the ATLAS or CMS photon ID criteria given that they pass the {\tt PGS4} photon ID criteria and the photon kinematic cuts, and normalize all efficiencies for diphoton searches computed using our MC simulation by this constant:
\be
\epsilon^{\gamma \gamma}_{BSM} \approx \epsilon^{\gamma \gamma \text{, MC}}_{BSM} \times \kappa
\ee
We chose to normalize to the $Vh$ production mode because the Higgs has somewhat similar kinematics to the BSM production modes we are considering. For comparison, applying the same procedure to the $gg \to h$ production mode we obtain $\kappa_{gg\to h}\sim0.6$.

The rates from the search due to SM Higgs production ($R_{\SM}$) and SUSY Higgs production in our model ($R_{\BSM}$) are thus given by
\begin{align}
R_{\SM}& = \sigma^h_{\SM} \times \epsilon_{\SM} \\
\Delta R_{\BSM}& = 2 \sigma_{\tilde{b}\tilde{b}} \times \epsilon_{\BSM} = 2\kappa\sigma_{\tilde{b}\tilde{b}} \times \epsilon_{\BSM}^{\text{MC}}
\end{align}
where we assume the experimentally reported efficiencies and theoretically predicted cross section for a 125 GeV Higgs to determine the SM contribution. Note that we define $\epsilon$ to include the branching ratio for the $h\to\gamma\gamma$ decay.

\begin{figure}
		  \centering
        \subfigure[]{
                \includegraphics[width=3in]{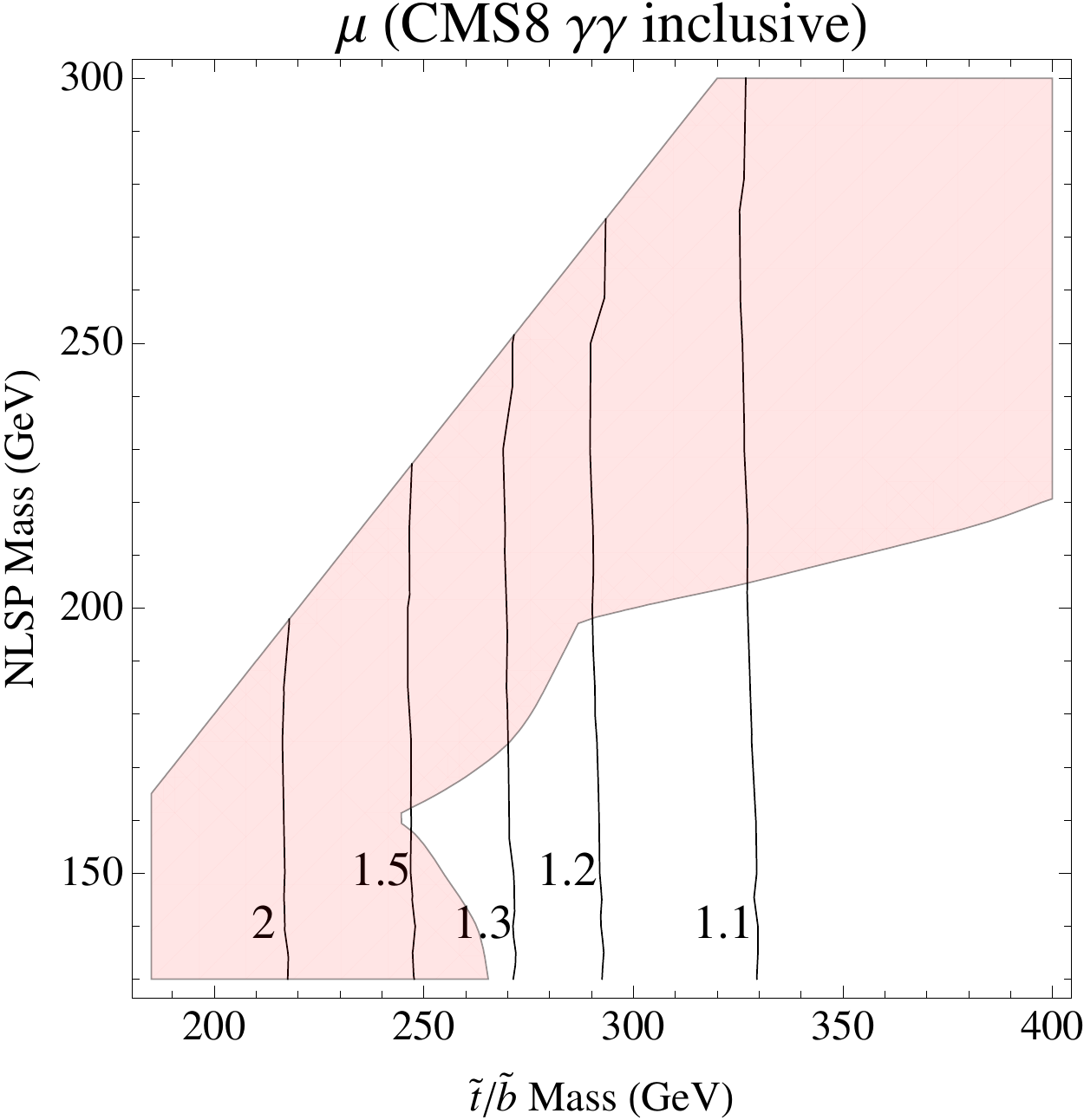}
                \label{fig:ggInclusive2D}
		  }
               \hspace{5mm} 
 	\subfigure[]{
                \includegraphics[width=3in]{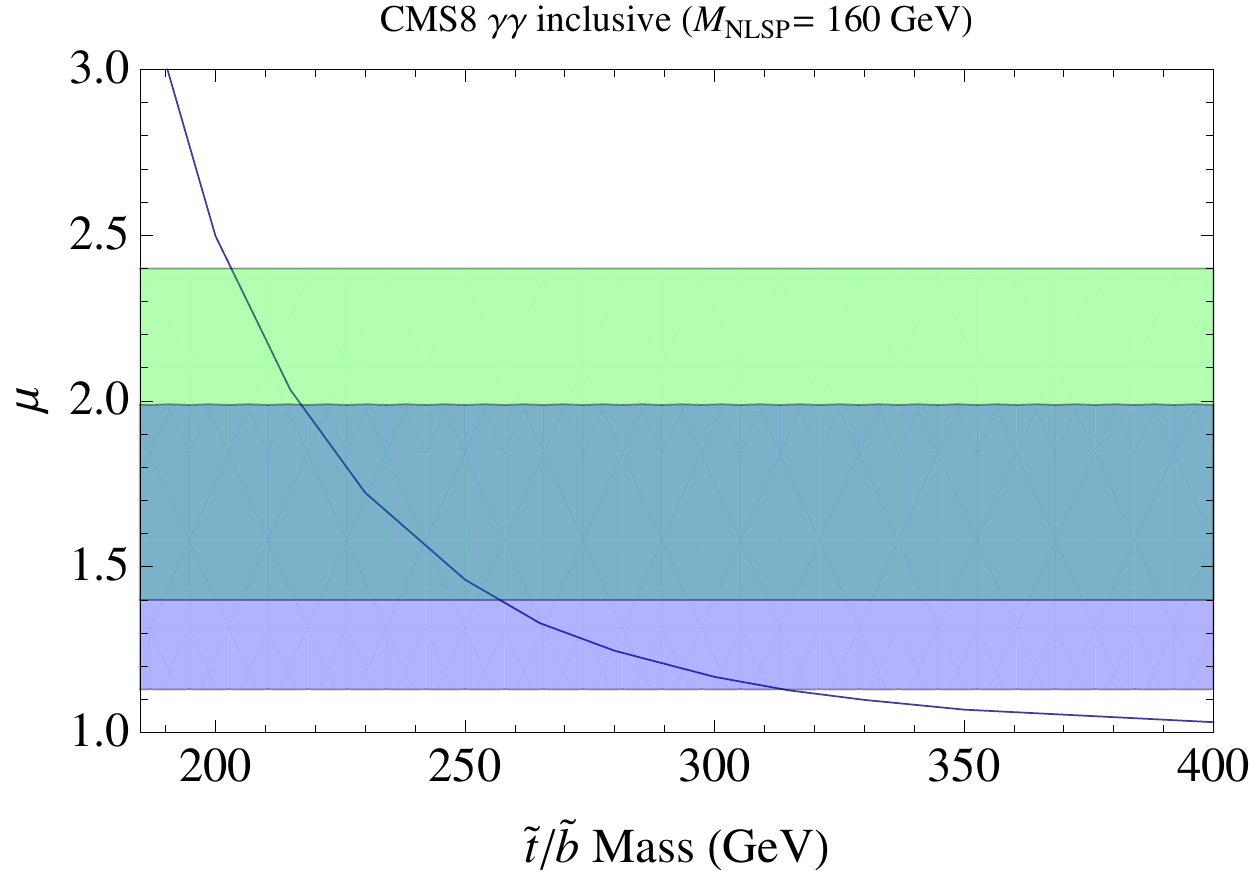}
                \label{fig:ggInclusive1D}
		  }

        \caption{(a):  $\mu$, the ratio of total BSM + SM rate to the SM rate, for the CMS8 inclusive diphoton search, as a function of squark mass and NLSP mass in the simplified model.  The shaded region indicates the existing 95\% exclusion at $\sim5\fb^{-1}$ from SUSY searches and the diphoton + MET search, shown in detail in fig. \ref{fig:CombinedLimits}.  (b): The same quantity as a function of squark mass only, for a NLSP mass of 160 GeV. Shaded in green and blue are respectively the ATLAS and CMS $1\sigma$ preferred regions. \label{fig:ggInclusive}}
\end{figure}

Using the above method we can (approximately) determine the contribution of Higgs production in our simplified model to the total rate in the ATLAS and CMS inclusive $h \to \gamma \gamma$ searches. We define $\mu$ to be the ratio of the total signal rate including BSM production to the SM rate,
$\mu = \frac{R_{\SM} + \Delta R_{\BSM}}{R_{\SM}}$. Fig.~\ref{fig:ggInclusive} shows $\mu$  as a function of the squark mass and NLSP mass in the 8 TeV CMS inclusive diphoton search. The efficiencies are insensitive to the stop/sbottom and NLSP mass, and $\mu$ therefore depends dominantly on the squark production cross section.  Because the squark cross section increases more rapidly with $\sqrt{s}$ then the Higgs cross section, at 7 TeV $\mu$ is about $10\%$ smaller. The rates at ATLAS are similar to those at CMS. For a 250 GeV stop/sbottom, the total rate is $\sim 150\%$ of the SM rate. We will discuss the possibility of explaining current hints of excesses in the inclusive diphoton channel in  section~\ref{sec:OtherSearch}.

\subsection{Diphoton plus lepton search} \label{sec:GGLepton}

CMS has published searches for diphoton resonances accompanied by a lepton or missing transverse energy (MET), targeting associated $hW$ and $hZ$ production with a focus on the fermiophobic Higgs model~\cite{FermiophobicCMS8}. Both of these final states are promising search channels for observation of BSM Higgs production. 

In our simplified model, the signal in the $\gamma\gamma+l$ channel occurs when one Higgs decays to photons while the other produces leptons through decays to $WW^*$, $\tau \tau$ or $ZZ^*$. Though the branching ratio to this state is small, the Standard Model background rate from non-Higgs processes and signal rate from $hZ\to \gamma\gamma\ell^+\ell^-$ or $hW\to\gamma\gamma\ell\nu$ production are both quite low.

\begin{figure}
		  \centering
                \includegraphics[width=3in]{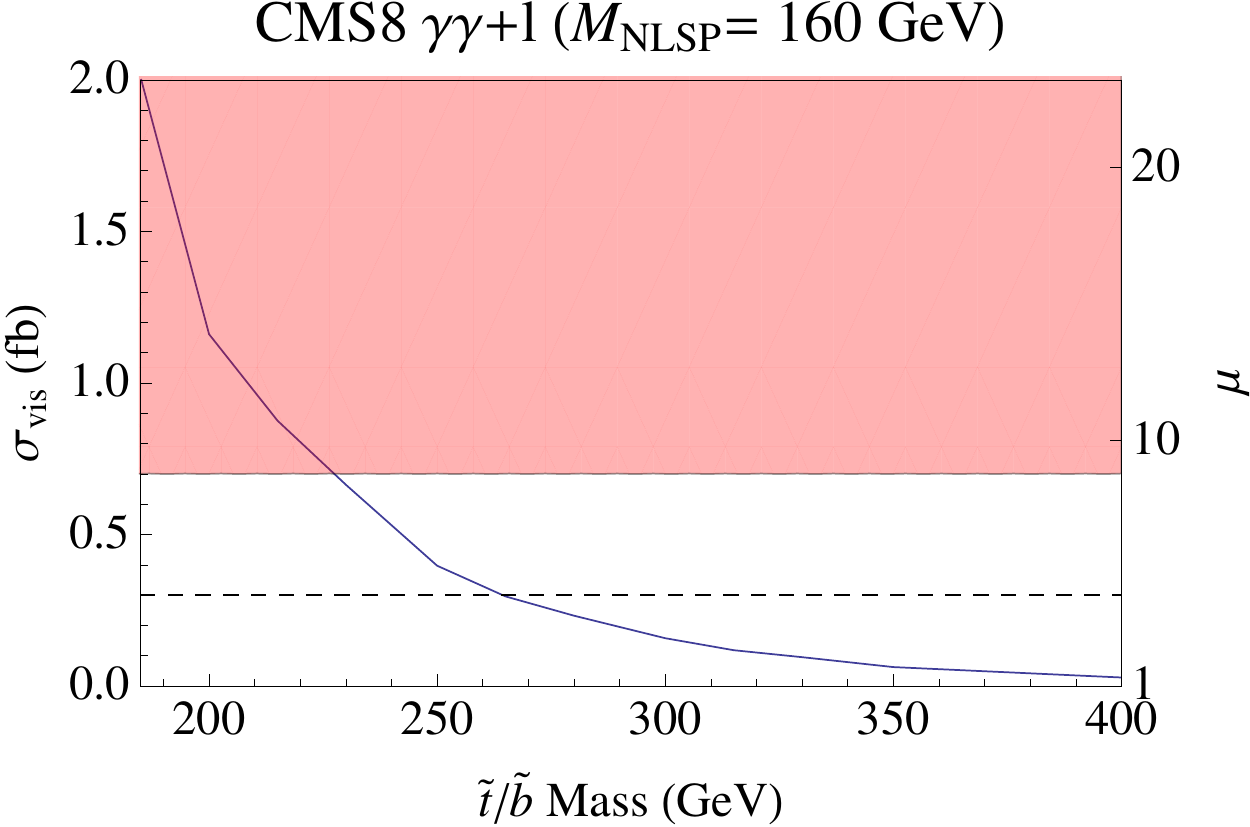}
              \caption{The rate from our simplified model in the CMS diphoton plus lepton search~\cite{FermiophobicCMS8}, both in events per fb and as a ratio $\mu$ to the rate from Standard Model Higgs production, as a function of squark mass for a NLSP mass of 160 GeV. The red shaded region is ruled out at 95\% CL by the existing $5.3\fb^{-1}$ of data at 8 TeV. Dashed is the \emph{expected} limit in $20\fb^{-1}$. }
\label{fig:LeptonDiphotonRate}
\end{figure}

As before, we calculate the contribution to the rate in this search from our model, adjusting our MC efficiencies by $\kappa_\CMS$, and plot the results against squark mass. Fig.~\ref{fig:LeptonDiphotonRate} shows the total rate both in units of cross section and as a ratio to the Standard Model rate. The search in $5.3 \fb^{-1}$ of 8 TeV data~\cite{FermiophobicCMS8} observed one event with about 1.6 expected background events within $125\pm 2$ GeV. With 0.5 SM Higgs events expected, this gives a 95\% confidence level bound of $0.7\fb$ on the visible cross section from new physics, excluding squark masses $\lesssim 240 \GeV$. Because the background is directly measured from the continuum diphoton spectrum, we scale the background uncertainty by $\sqrt{L}$ to obtain an expected limit of $0.3\fb$ in 20$\fb^{-1}$.

We validated our results by comparing our computed efficiencies for SM $Wh$ and $Zh$ events to the efficiencies determined from \cite{FermiophobicCMS8} for a 120 GeV fermiophobic Higgs, finding agreement within 10\%.

\subsection{Diphoton plus missing energy search}\label{sec:GGMET}

We now turn to the CMS search for diphoton + missing transverse energy (MET) targeting $Zh \to \nu\nu\gamma\gamma$ ~\cite{FermiophobicCMS8}. The Standard Model signal in this search is suppressed by branching ratios and the small $Zh$ production cross section. However, this search can be quite sensitive to Higgs production in NLSP decays due to the guaranteed MET from the LSP, as long as the NLSP and Higgs are not too degenerate. 

\begin{figure}

 		\includegraphics[width=3in]{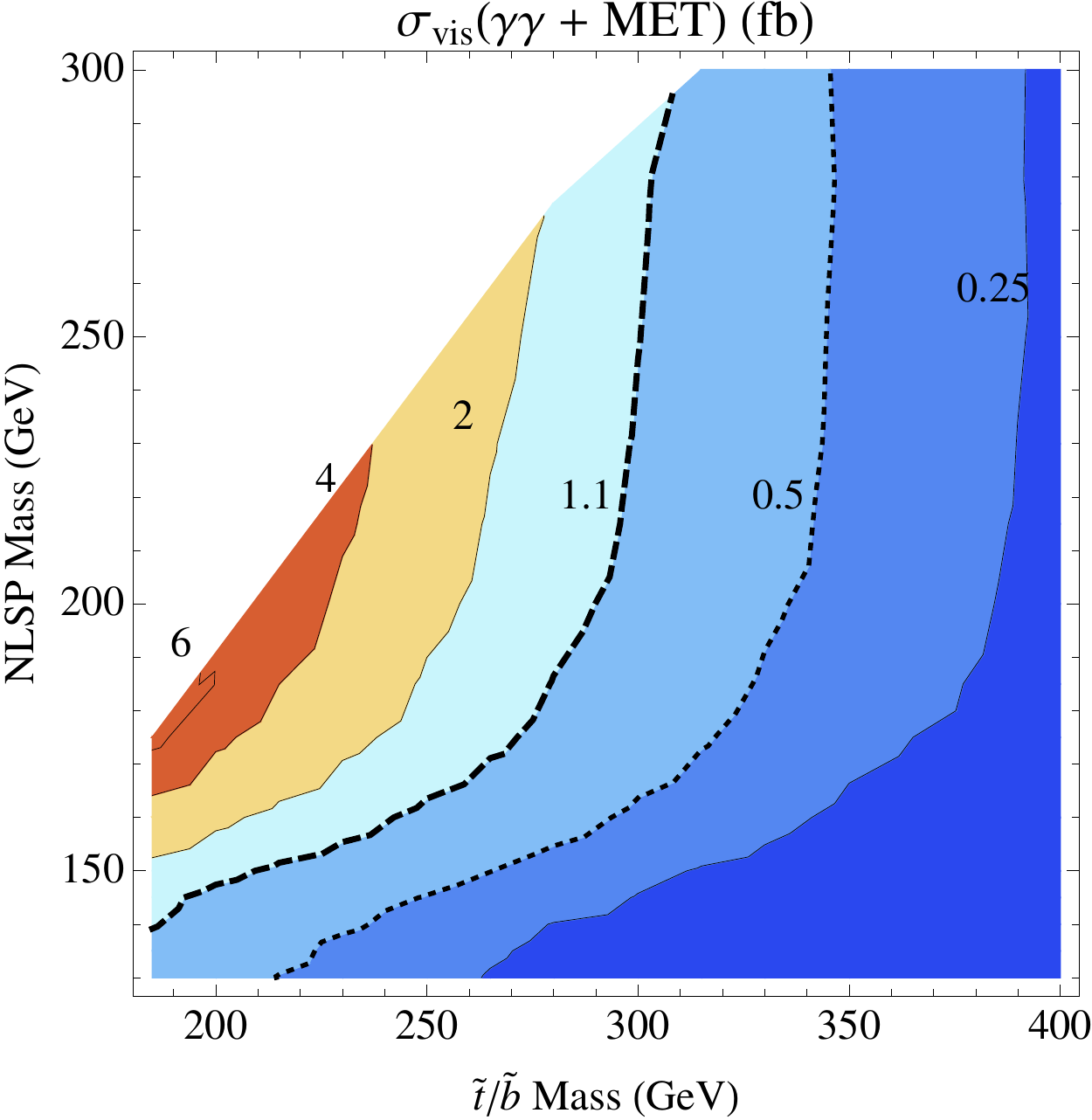}
           
               \hspace{5mm}

              \caption{The visible cross section for our simplified model in the CMS diphoton plus MET search~\cite{LeptonDiphoton}  as a function of squark mass and NLSP masses. We take the $95\%$ limit from $5.3\fb^{-1}$ of data at 8 TeV to be 1.1 fb.}
\label{fig:GGMET702D}
\end{figure}

The kinematic selections on the photons for this search are similar to the other diphoton channels, except that only photons within the barrel are selected. Additionally, events passing the lepton tag are vetoed. Events with missing transverse energy greater than 70 GeV are selected. The background rate (fit from data) was $2.4\pm0.8$ events in the range $125\pm 2$ GeV, and 3 events were observed in $5.3 \fb^{-1}$. With a SM higgs expectation of 0.2 events, this gives a 95\% limit on the visible cross section from new physics of 1.1 fb. The expected visible cross section for this search in the simplified model is shown as a function of squark and NLSP mass in fig.~\ref{fig:GGMET702D}. In fig.~\ref{fig:CombinedLimits}, the 95\% exclusion are compared to the limits from SUSY searches, both for current data and extrapolated to $20\fb^{-1}$.  Since the background in this channel is fit from continuum diphoton data, we scaled the systematic uncertainty as $\sqrt{L}$ to estimate the expected limits in $20\fb^{-1}$.  This search has better reach than the SUSY searches for NLSP masses of $\sim170-200\GeV$.

For NLSPs lighter than $\sim 150 \GeV$, the signal in our model is suppressed significantly by a  cut of MET $>$ 70 GeV, which removes 80-90\% of the signal events passing the other selections. The sensitivity of such a search can therefore likely be improved by including one or more additional channels with lower MET threshold, as was done in the non-resonant CMS diphoton + MET search \cite{CMSdiphotonMETnonresonant}. As an example, the bin $40\GeV < \MET < 70 \GeV$ generally contains 2-3 times as many signal events as the $\MET>70 \GeV$ bin. If the background scales with the MET cut as in the non-resonant search~\cite{CMSdiphotonMETnonresonant}, it would be a factor of about 20 times larger in this bin, giving $S/\sqrt{B}$ about half that of the higher MET channel. As the background increases rapidly with lower MET cuts, sub-dividing this region could further increase the sensitivity in lower MET channels. 

As before, we validated our results against the experimental efficiencies for a 120 GeV fermiophobic Higgs.  Our analysis yields a 125 GeV SM Higgs signal rate of $R^{\gamma \gamma+MET}_\SM=.04 \fb$, which agrees within 10\%  with the efficiency from the expected signal rate in~\cite{FermiophobicCMS8}.

\subsection{Discovering SUSY in the $(h\to\gamma\gamma)+1/2b$ channel}\label{sec:ggb}

The Standard Model background rate for the $h\to\gamma\gamma$ search in $5.3\fb^{-1}$ at $8\TeV$ at CMS is 433.8 events/GeV near the Higgs mass peak~\cite{DiphotonCMS}, and consists dominantly of prompt diphoton ($\gamma\gamma$) events and single prompt photon events with one or more jets ($\gamma j$), with one jet faking a photon~\cite{CMSdiphotonBGMC}. This background can be considerably reduced by requiring one or two $b$-tags, while maintaining very good efficiency for the diphoton signal from our simplified model, which produces copious amounts of $b$ quarks in the final state. We therefore consider a search with the same cuts as the CMS diphoton search, but with the additional requirement of one or two $b$-tagged jets with $p_T>50 \GeV$. We consider a counting experiment in which the number of events with diphoton mass within 2 GeV of $m_h = 125 \GeV$ is compared to the background expectation. The sensitivity of a similar search has been studied in refs.~\cite{Baur:2003gp, Contino:2012xk} in the context of measuring variations in the Higgs couplings in Higgs pair production events at very high luminosities with the 14 TeV LHC, and the sensitivity of a diphoton + jets search at 14 TeV without $b$-tags has been considered in \cite{Azatov:2012rj} for studying fermionic top-partner decays to the final states $th+X$.

To estimate the continuum, non-Higgs background from $\gamma\gamma$ and $\gamma j$, we simulated $\sqrt{S}=8\TeV$ MLM matched samples of $\gamma\gamma +0/1/2j$ and $\gamma+1j/2j/3j$, where $j$ is a light flavor or gluon, a charm quark, or $b$ quark at 8 TeV. As before we simulate the response of the CMS detector with the default PGS CMS parameter card, but here we use a different modified $b$-tagging function which gives a flat tag rate of $1\%$ for light flavor/gluon jets, $15\%$ for charm, and $60\%$ for $b$ quarks \cite{CMSbtag}. We separately normalize the $\gamma\gamma$ and $\gamma j$ rates to the rate in \cite{DiphotonCMS}  with relative proportions of 25\% $\gamma j$ and 75\% $\gamma\gamma$ based on the Monte Carlo results in \cite{CMSdiphotonBGMC}. The results for the background rates are shown in table \ref{tab:ggbRates}. The first $b$-tag reduces the non-Higgs background by a factor of 100 to $\sigma_{\text{cont.},1b}=2.5\fb$, with the remaining contributions from true $b$-jets, light quark fakes, and charm fakes all roughly comparable. The second $b$-tag reduces this background by another factor of 50 to $\sigma_{\text{cont.},2b}=0.05\fb$, with the remaining contributions dominantly from true $b$-jets or mistagged charm.

\begin{table}
\begin{tabular}{c||ccc|c||ccc|c||c}
 				& $\gamma\gamma$ 	& $\gamma j$	& $t\overline{t}$ & non-$h$ tot. & $t\overline{t}h$	& $b\overline{b}h$	& $Zh$ 	&SM $h$ tot. 	& signal \\
\hline
$\sigma(8\TeV) $ [fb]		&	-				&	-	& $230\times10^3$ & -		& $130$			&   $ 200$		& $394$		& -		&	$2\times10^3$ \\
$2\gamma, m_{\gamma\gamma}=m_H\pm2\GeV$ [fb] & 
							250				&	80		& 0.17		& 330 	&	0.1			&	0.2			& 0.2		& - 		 &	1.8\\
$1b,\;p_T>50\GeV $ [fb]		&	1.8				&	0.6		& 0.08		& 2.5	&	0.05			&	0.01			& 0.01		& 0.07	 &	1.4\\
$2b,\;p_T>50\GeV$  [fb]		&	0.025			&     0.007 	& 0.02		&0.05   	& 	0.02			&	0.0004		& 0.0009		& 0.02	 &	0.7
\end{tabular}
\caption{
The visible cross section for non-Higgs continuum and SM Higgs diphoton rates after various cuts in the $\gamma\gamma+1/2b$ search, including all detector efficiencies, branching ratios, and fake rates. The $\gamma\gamma$ and $\gamma j$ categories include the non-resonant, non-electroweak contributions to the rate in events with 2 real photons or 1 real photon and 1 jet faking a photon respectively. $t\overline{t}$ is the contribution from top quark pair production with hard radiated photons or electrons/jets faking photons. $t\overline{t}h$ is a Higgs-strahlung process in the SM, while $b\overline{b}h$ includes this and also QCD production of a $b\overline{b}$ pair in gluon fusion Higgs production. For $b\bar{b}h$ we use the tree level cross section from {\tt MadGraph5}. $hZ$ is the associated production of a SM Higgs and $Z \to b \bar{b}$. The cross section for $t\overline{t}$ is from \cite{TopXSCMS} and for $t\bar{t}h$ and $hZ$ from \cite{LHCHiggsCrossSectionWorkingGroup:2011ti}.  The $\gamma\gamma$ and $\gamma j$ contributions are normalized to the observed background in \cite{DiphotonCMS} as described in the text. The signal rate in the simplified model for a 300 GeV squark and 150 GeV NLSP is also shown.
\label{tab:ggbRates}
}
\end{table}

For the $2b$ selection, top pair production with hard radiated photons or electrons/jets faking photons is also an important continuum background. To estimate this, we take a jet-photon fake rate of $0.1\%$ \cite{PhotonPerformanceATLAS} and an electron-photon fake rate of $2\%$ \cite{CMSdiphotonMETnonresonant}. The estimated rate is $\sigma_{t\overline{t},2b}=0.02\fb$, with fake electrons and radiated photons giving the dominant contributions. SM $t\bar{t}h$ also gives a resonant contribution to the background with a rate comparable to the continuum background. $Zh$ and $b\bar{b}h$ production give a negligible rate, as does non-resonant electroweak production with prompt or fake photons. The 2$b$ channel is therefore essentially backgroundless at $5\fb^{-1}$ and in  $20\fb^{-1}$ has $\sim 1$ expected continuum event falling in the Higgs mass window and  $\lesssim 1$ expected SM $t\bar{t}h$ event. Importantly, because this is a search for a known resonance, the background rate can be measured from data as in the usual Higgs diphoton search, so an actual search would be limited only by statistical uncertainties in the background fit.

\begin{figure}
	  \centering
        \subfigure[]{
 		\includegraphics[width=3in]{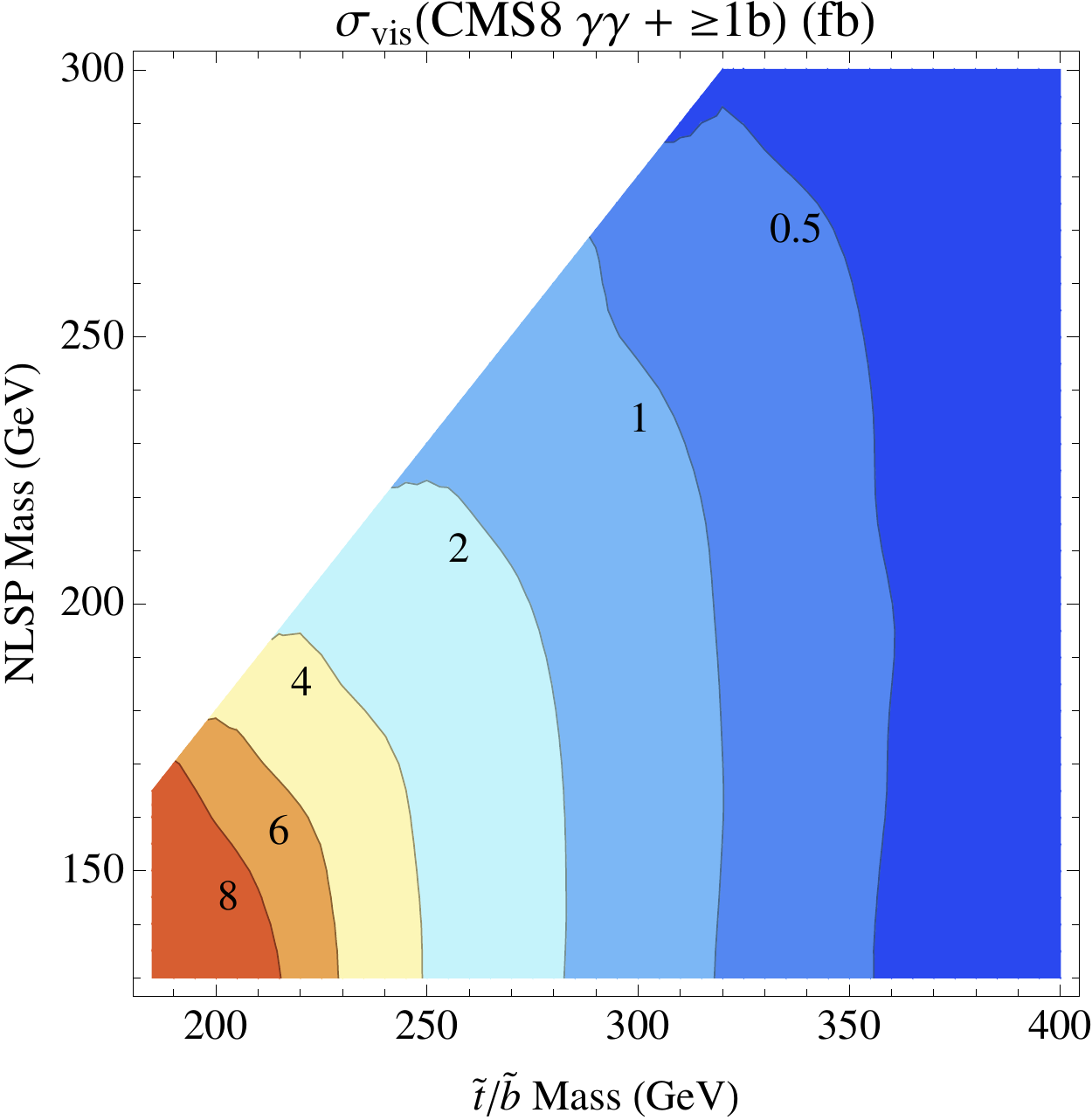}
		  }
               \hspace{5mm} 
 	\subfigure[]{
		 \includegraphics[width=3in]{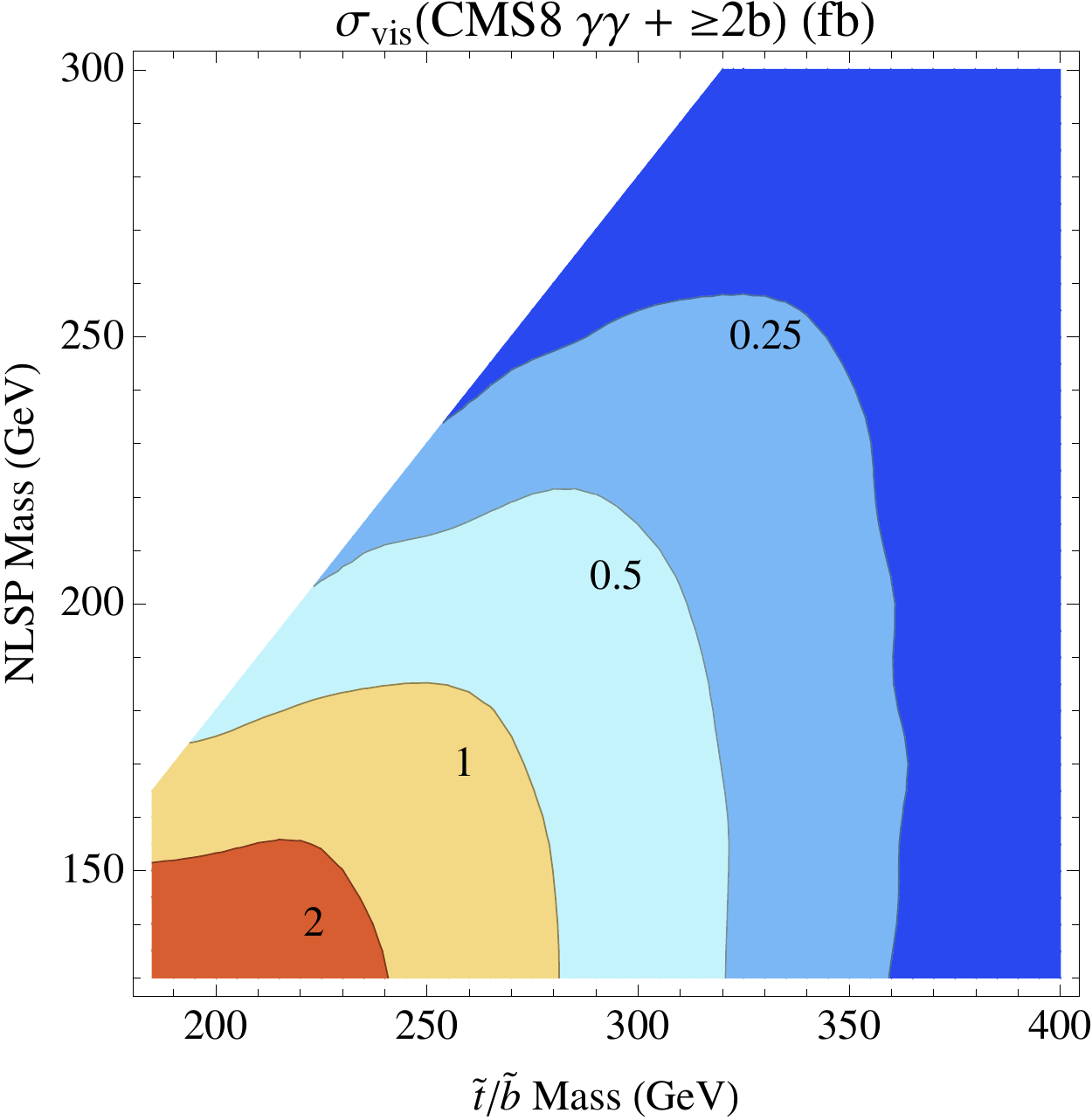}
		  }
		  \centering
               
              \caption{The visible cross section for our simplified model in a search for $h\to\gamma\gamma$ in association with $b$-tagged jets.  (a) Requiring at least one $b$-tagged jet with $p_T>50\GeV$.  (b) Requiring at least two $b$-tagged jets with $p_T>50\GeV$ \label{fig:bggVisXS}.}
\end{figure}

The visible cross section for the $1b$ and $2b$ selections are shown for our simplified model in fig. \ref{fig:bggVisXS}. The efficiency for signal events passing the inclusive diphoton selections to also pass the $1b$ selection is $50-70\%$, while for the $2b$ selection it is $20-40\%$. For the $1b$ channel we estimate the systematic uncertainty in a background fit to be $20\%$ and $10\%$ at $5\fb^{-1}$ and $20\fb^{-1}$ respectively based on counting statistics in the neighboring diphoton invariant mass bins. With a background rate of $\sigma_{\text{BG},1b} = 2.5\fb$, we find expected $95\%~\text{CL}_{\text{s}}$ limits on the visible cross of $1.9\fb$ for $5\fb^{-1}$ and $0.9\fb^{-1}$ for $20\fb^{-1}$. The $1b$ channel has sensitivity significantly exceeding the other diphoton and SUSY searches we have discussed, as shown in fig. \ref{fig:CombinedLimits}.

\begin{figure}
	  \centering
        \subfigure[]{
 		\includegraphics[width=3in]{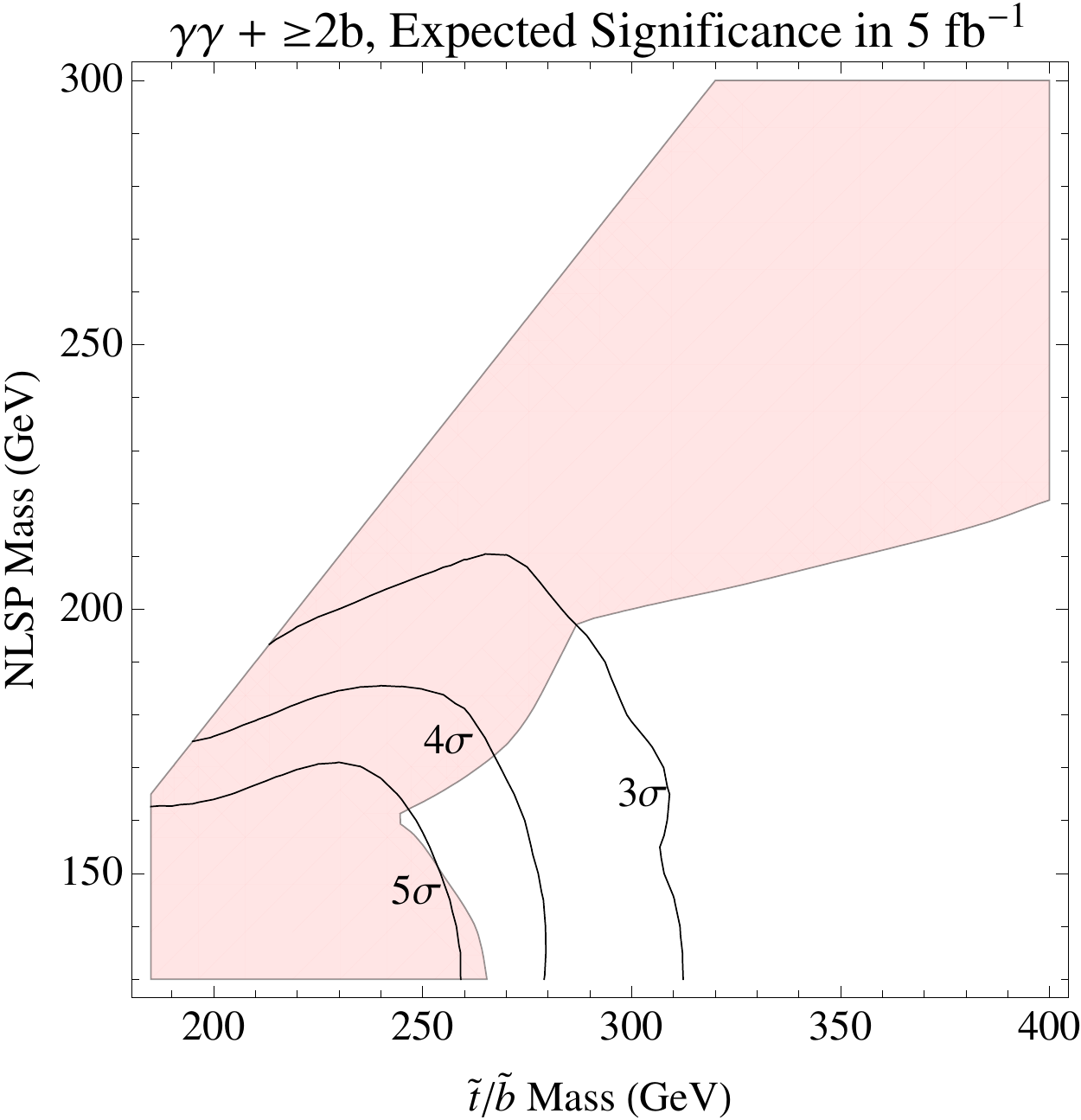}
		  }
               \hspace{5mm} 
 	\subfigure[]{
		 \includegraphics[width=3in]{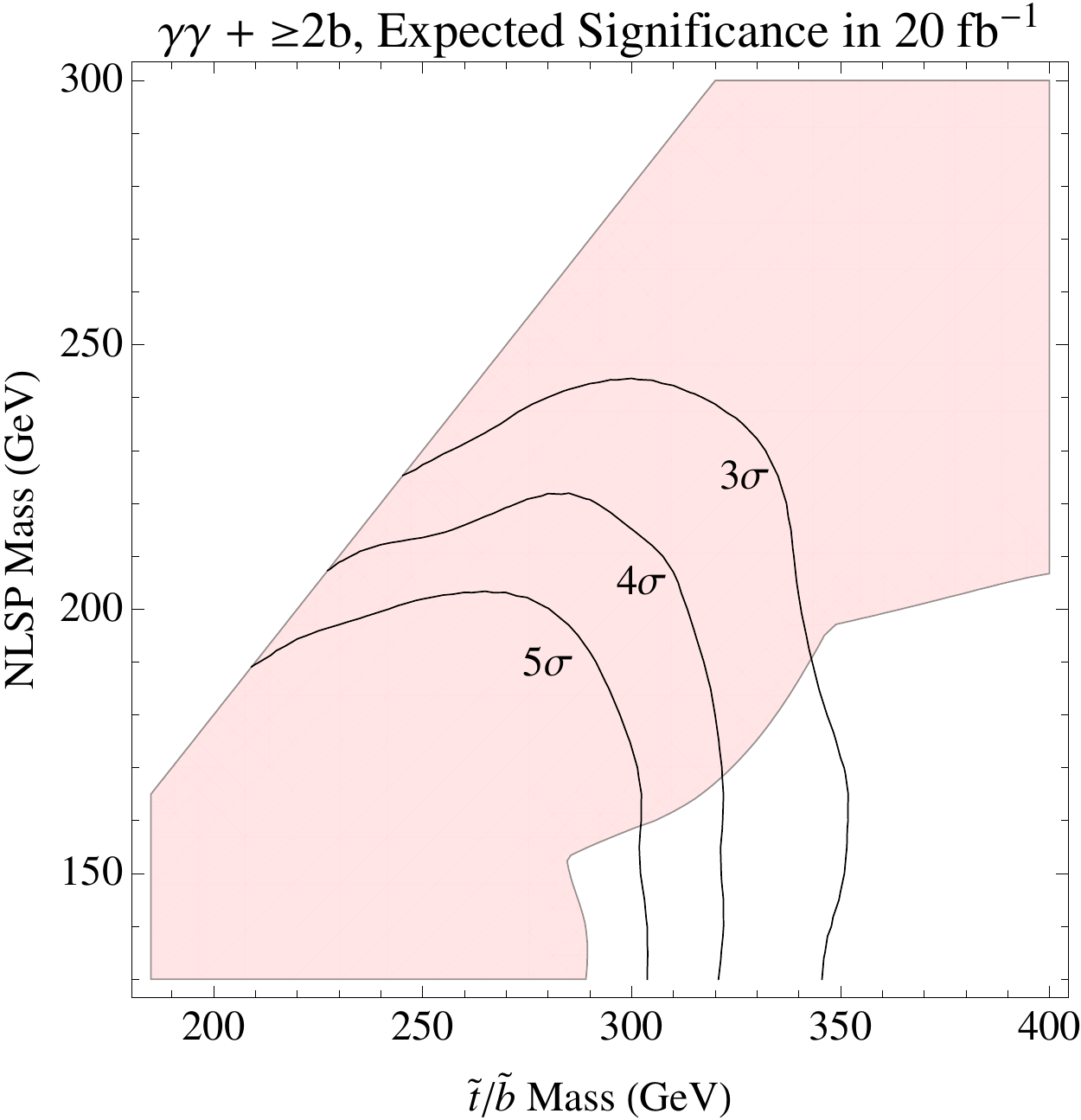}
		  }
		  \centering
               
              \caption{The expected signal significance for a proposed search in $\gamma\gamma+2b$ in 8 TeV with (a) $5\fb^{-1}$ and (b) $20\fb^{-1}$. The shaded region indicates (a) the existing 95\% exclusions at $\sim5\fb^{-1}$ and (b) the extrapolated limits at $20\fb^{-1}$ from SUSY searches and the diphoton + MET search, as described in fig. \ref{fig:CombinedLimits}.\label{fig:2bdiscovery}}
\end{figure}

The near backgroundless $2b$ channel is even more sensitive to our model and can even lead to a discovery at the 8 TeV LHC. Since there are large uncertainties in our background estimate due to the fake rates for photons and $b$-tags, we estimate the reach of the $2b$ channel taking the background rate to be $0.1\pm0.05 \fb$.  Fig.~\ref{fig:2bdiscovery} shows the expected signal significance in $5$ and $20\fb^{-1}$, along with the 95\% CL excluded region from the existing SUSY and diphoton+MET searches and their extrapolated limits in 20$\fb^{-1}$. For direct comparison with the other searches, the expected 95\% exclusions from this channel are also shown in fig. \ref{fig:CombinedLimits}. 

In 5$\fb^{-1}$ the $\gamma\gamma+2b$ search already has significantly better reach than the other searches we have considered, and in $20\fb^{-1}$ it gives an expected significance of $>3\sigma$ for squarks up to $350\GeV$, well beyond even the exclusion reach for the other searches, and a potential  $5\sigma$ discovery for squarks up to $300\GeV$. Another important advantage of this channel over other SUSY searches that can probe our models is that by exploiting the Higgs diphoton resonance, this search clearly identifies the the involvement of an on-shell Higgs boson in the new physics. In contrast, a signal observed in the multi-$b$ or leptonic searches would not immediately suggest excess Higgs production.

\section{Diphoton excesses and other LHC Higgs searches} \label{sec:OtherSearch}

Ref. \cite{Giardino:2012dp} presents a combined fit to the ATLAS and CMS 7 and 8 TeV signals in the various SM Higgs searches. The only $>1\sigma$ deviations from the SM predictions are excesses in the inclusive and VBF tagged $\gamma\gamma$ channels and a low rate in the $\tau\tau$ channel, driven primarily by a low rate in the VBF tagged $h\to\tau\tau$ channel at CMS \cite{CMSDiscovery}.

As shown in fig. \ref{fig:ggInclusive}, for a squark mass of 250-290 GeV, an enhancement of 20-50\% in the diphoton channel from Higgs production in SUSY cascades is possible while remaining consistent with other searches. This search is sensitive to our model because it makes use of the clean $\gamma\gamma$ resonance and does not place cuts on the non-Higgs activity in the event. This is also true for the $h \to ZZ^* \to 4\ell$ resonance searches, so we would expect any enhancement in the inclusive diphoton searches by this mechanism to give a comparable enhancement to $h \to ZZ^* \to 4\ell$.

On the other hand, searches in other channels often require additional cuts on the event to reduce SM backgrounds and consider more complicated kinematic variables, resulting in lowered signal strength from the non-standard Higgs production modes we consider. For example, in order to reduce backgrounds from top production, the ATLAS and CMS searches in the $h \to WW^* \to l\nu l\nu$~\cite{ATLASWW7,CMSWW7} and in most channels for $h \to \tau \tau$~\cite{ATLAStautau7,CMStautau7} veto events with any $b$-tagged jets. Since events from our models typically contain several $b$-jets, they give a negligible contribution to the rate in these searches. Furthermore, the kinematic variables considered in these analyses (such as the transverse mass of leptons and the reconstructed tau-tau invariant mass) are distorted by the different kinematics and additional missing energy in events from squark production, so those events that are selected will not necessarily increase the signal strength for the 125 GeV SM Higgs hypothesis. Similar issues limit the sensitivity of searches for $Vh$ production with $h \to b \bar{b}$ and the vector boson $V$ decaying to leptons and/or neutrinos~\cite{ATLASVbb7,CMSVbb7}. Although our model produces these same final states, the $b$-jets selected by the analyses will not necessarily be the Higgs decay products, so a mass peak will not be visible. Both analyses place many kinematic cuts on events, such as requiring the $b\bar{b}$ momentum to be nearly back-to-back with the reconstructed vector boson momentum when projected into the transverse plane, greatly reducing the efficiency in our case. 

In the $h \to \gamma \gamma$ analyses both CMS and ATLAS define dijet-tagged event classes, selecting jets widely separated in pseudorapidity $\eta$ to obtain sensitivity to VBF Higgs production in the SM. The additional jets present in BSM events from our simplified model rarely satisfy this $\Delta\eta$ requirement, and the enhancement relative to the SM rate is at most a few percent in the parameter range we consider.

To summarize, if the simplified model of excess Higgs production we have considered gives an observable enhancement to inclusive $h\to\gamma\gamma$ rate, then this is consistent with the lack of observed excesses in $h \to WW^*$, $h \to \tau \tau$, and $Vh \to Vb\bar{b}$, but is in slight tension with an observed rate for $h\to 4\ell$ which is consistent with the SM and does not explain the excess in the dijet-tagged $h \to \gamma \gamma$ channel. Therefore, while the enhancement to the inclusive diphoton channel is encouraging, the simplified model alone likely does not give a significantly better fit to current data than does the SM prediction.

Thus far we have always assumed that the effective couplings of the Higgs are the same as those predicted by the Standard Model. However, the light third generation squarks of our models can modify the Higgs couplings to gluons and photons through loop effects, particularly if there are light stops (discussed recently in refs. \cite{Arvanitaki:2011ck,Blum:2012ii,Giardino:2012ww,Buckley:2012em}). Large stop mixing (e.g. from $A$-terms) is required to give the correct sign to enhance the diphoton branching ratios, and extremely large mixing is required to actually enhance the total diphoton rate from gluon fusion Higgs production since the $gg\to h$ rate is at first suppressed as the diphoton branching ratio is enhanced~\cite{Giardino:2012ww}. By contributing another production mode which is not sensitive to the $gg\to h$ coupling, Higgs production in cascade decays allows an overall enhancement to the diphoton rate at smaller mixings. The modified $h\to\gamma\gamma$ branching ratio then also enhances the dijet-tagged diphoton rate to reproduce the observed excess, while the reduced $gg\to h$ rate prevents an excess in the $h \to ZZ^* \to 4l$ channel. 

We do not attempt a detailed fit to the data, but rather note that in the range of squark masses where these models can give significant enhancements to inclusive SM Higgs searches, it is likely that excesses in the $\gamma\gamma+\MET$ search and SUSY searches discussed in sec.~\ref{sec:GGMET} and~\ref{sec:Constraints} will be observed in 20$\fb^{-1}$ of 8 TeV data and that a discovery could be made in a search for $\gamma\gamma+1/2b$ as described in sec.~\ref{sec:ggb}. As a general point, excess Higgs production in BSM cascades, possibly with less dramatic exclusive signatures than in our models, is an interesting degree of freedom to consider when fitting rates in Higgs searches, giving qualitatively different effects from modifications to Higgs couplings. 

We also note that to improve sensitivity,  both ATLAS and CMS divide the inclusive diphoton search into several event classes based on the kinematics of the photon and the quality of its reconstruction. Our results in sec. \ref{sec:GG} based on the inclusive diphoton count therefore only give an approximation of expected best fit rates, which in refs. \cite{DiphotonATLAS,DiphotonCMS} are actually determined with a fit taking into account the different signal and background expectations in the individual event classes. Of particular note, both ATLAS and CMS have event classes capturing Higgs production with higher transverse momentum, which has a smaller background rate. For example, ATLAS separates diphoton candidates into high and low $p_{Tt}$ event classes, where $p_{Tt}$ is the component of the diphoton momentum transverse to the thrust axis~\cite{DiphotonATLAS}. The simplified model Higgs production leads to a substantially larger signal enhancement in the high $p_{Tt}$ category compared to SM $gg\to h$ events. For a given enhancement to the inclusive rate relative to the SM, the expected enhancement in the high $p_{Tt}$ categories relative to the SM rate in these channels is generally $4-5$ times greater. In the combined 7 + 8 TeV results these channel show roughly equal enhancements compared to the SM rate, but with large enough individual uncertainty to accommodate a substantially higher rate in the high $p_{Tt}$ channels. While other exclusive diphoton channels are more sensitive to our specific model, with more statistics this could be a useful constraint on more general possibilities for enhancing Higgs production through BSM cascades.

\section{Direct production of neutralinos and charginos} \label{sec:DirectProd}

If the NLSP in our models is a sufficiently light wino or higgsino, then the electroweak direct pair production rates can be large compared to SM associated Higgs production rates. This can mildly constrain the neutralino/chargino spectra possible for the colored production case, and can also enhance the signal in diphoton searches. As discussed in sec. \ref{sec:ChiToHiggs}, we consider the possibilities of wino co-NLSPs with small splittings such that the charginos decay promptly to the LSP, and a mixed wino-bino or higgsino NLSP, with the nearby neutralinos and charginos decaying promptly to the NLSP.

\begin{figure}
		  \centering
                \includegraphics[width=2.5in]{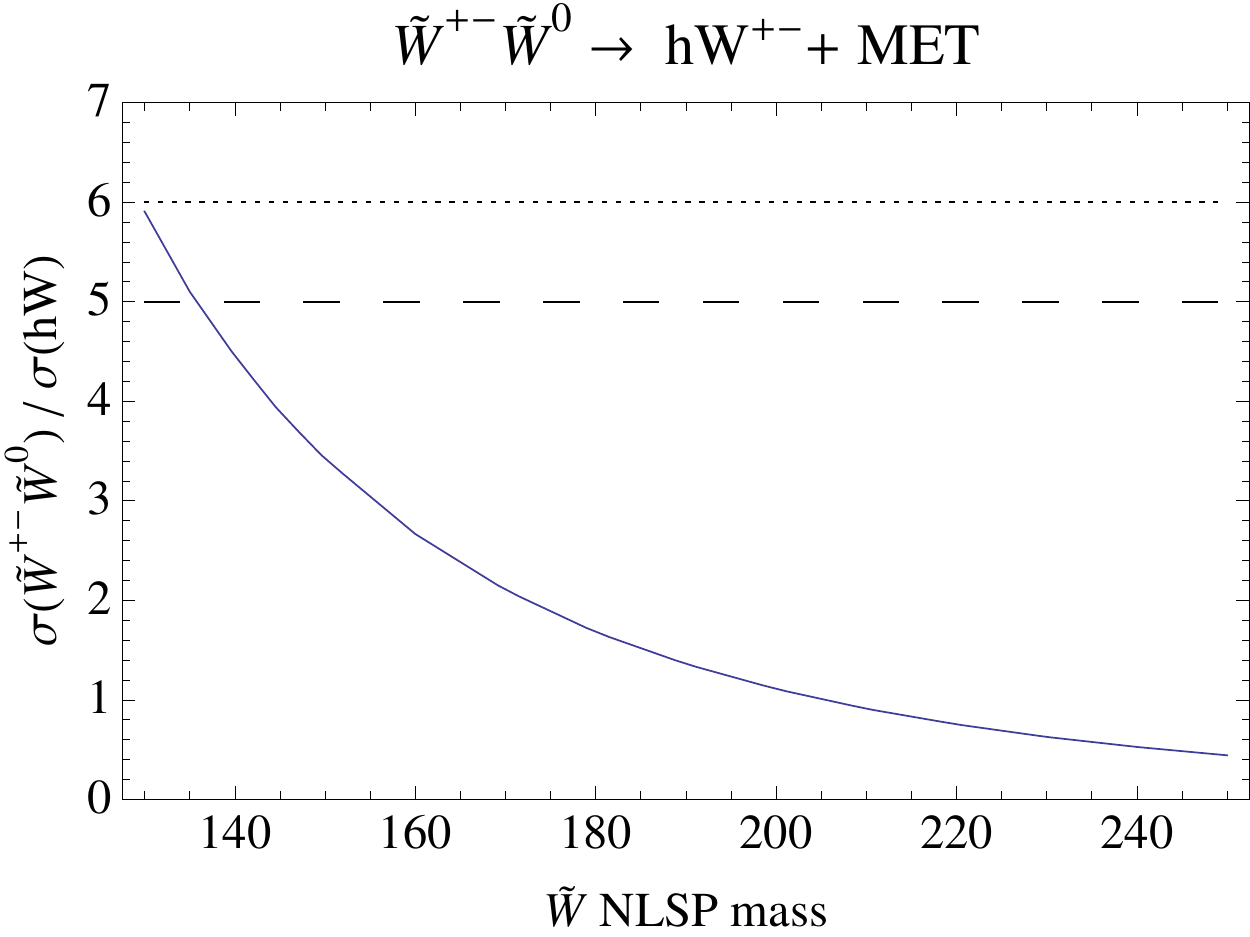}
              \caption{The cross section for wino co-NLSP direct production relative to the SM cross section for $hW$ associated production at 8 TeV. Dashed is the 95\% CL limit on the BSM contribution in the ATLAS $hW\to3l3\nu$ channel, and dotted is the limit from the CMS $\gamma\gamma+l$ channel. \label{fig:WinoDirectXS}}
\end{figure}

For wino-like co-NLSPs, the dominant production modes are $\chi_2^0 \chi_1^\pm$ and $\chi_1^+\chi_1^-$ (here $\chi_2^0$ denotes the neutral NLSP and $\chi_1^0$ the LSP). The winos decay as $\chi^\pm_1 \to W^\pm \chi_1$ and $\chi_2 \to h \chi_1$, so the final states of interest are $hW+\MET$ and $W^+W^-+\MET$.  Refs.~\cite{Lisanti:2011cj, Curtin:2012nn} discuss the possibility that $\chi_1^+\chi_1^-$ production can appear in $h\to W^+W^-$ search channels, but for $\chi_1^\pm$ decaying dominantly to $W^\pm$ there is no significant constraint on direct production in the parameter space of interest. The discovery potential for heavier charginos and neutralinos in the $hW$ mode at the 8 and 14 TeV LHC has been studied in refs.~\cite{Ghosh:2012mc,Baer:2012ts}. However for a lighter NLSP with small splitting with the Higgs, the MET signature is suppressed and the resulting signals become much more similar to SM associated Higgs production. Because the production cross section can be many times larger than the SM rate and there is no additional activity in the event (unlike in the colored production case), all of the existing searches for SM-like $hW$ associated production will directly constrain the $\chi_2^0\chi_1^\pm$ production cross section, which is shown relative to the SM $hW$ cross section at 8 TeV in fig.~\ref{fig:WinoDirectXS}.  The most constraining direct limit on $hW$ is an ATLAS search is for $hW \to 3l3v$ at 7 TeV with a limit of $\sigma_{Wh}/\sigma_{Wh,\text{SM}} \lesssim 6$~\cite{ATLAShW3l3v}, requiring $m_{\tilde W}\gtrsim135\GeV$; a similar bound in the same channel is also presented in ref. ~\cite{ATLAS3lDirect} in a search for direct production of gauginos with leptophilic decays. The limit from the CMS $\gamma\gamma+\ell$ search~\cite{FermiophobicCMS8} discussed earlier is $\sigma_{\text{WH}}/\sigma_{\text{WH,SM}} \lesssim 7$.  The strongest combined constraint on $Zh$/$Wh$ production currently comes from a CMS search for $h \to b\overline{b}$ with leptonically decaying $W$'s and $Z$'s at 7 and 8 TeV, which gives a 95\% upper limit of $\sigma_{\text{VH}}/\sigma_{\text{VH,SM}} < 2.11$ for a 125 GeV Higgs~\cite{CMShV}, but individual bounds on $hW$ and $hZ$ production were not presented. Ref. \cite{Byakti:2012qk} has suggested using jet substructure methods to further improve sensitivity for direct production of gauginos decaying in the $b\overline{b}W$ channel.

\begin{figure}
		  \centering
        \subfigure[]{
                \includegraphics[width=2.5in]{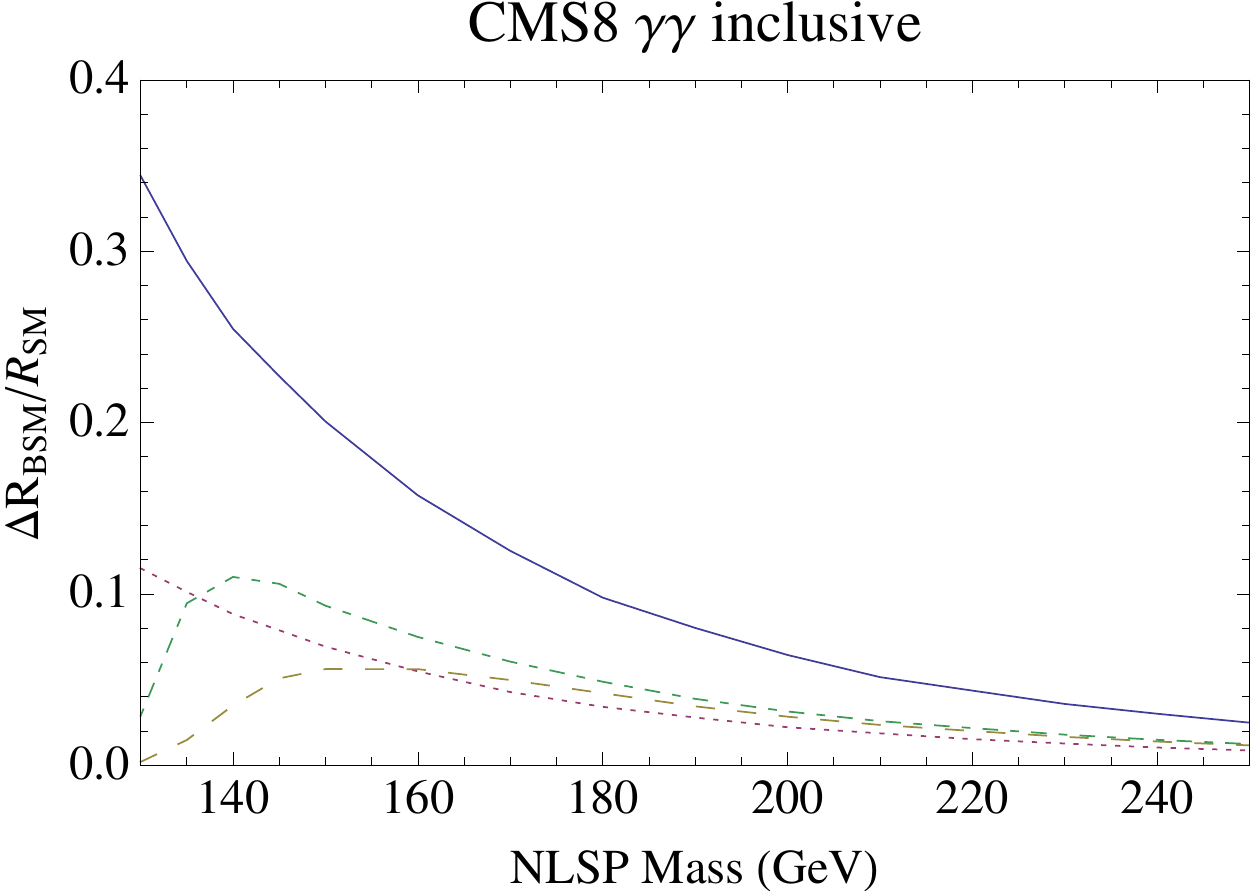}
                \label{fig:directDiphotonInc}
		  }
               \hspace{5mm} 
        \subfigure[]{
                \includegraphics[width=2.5in]{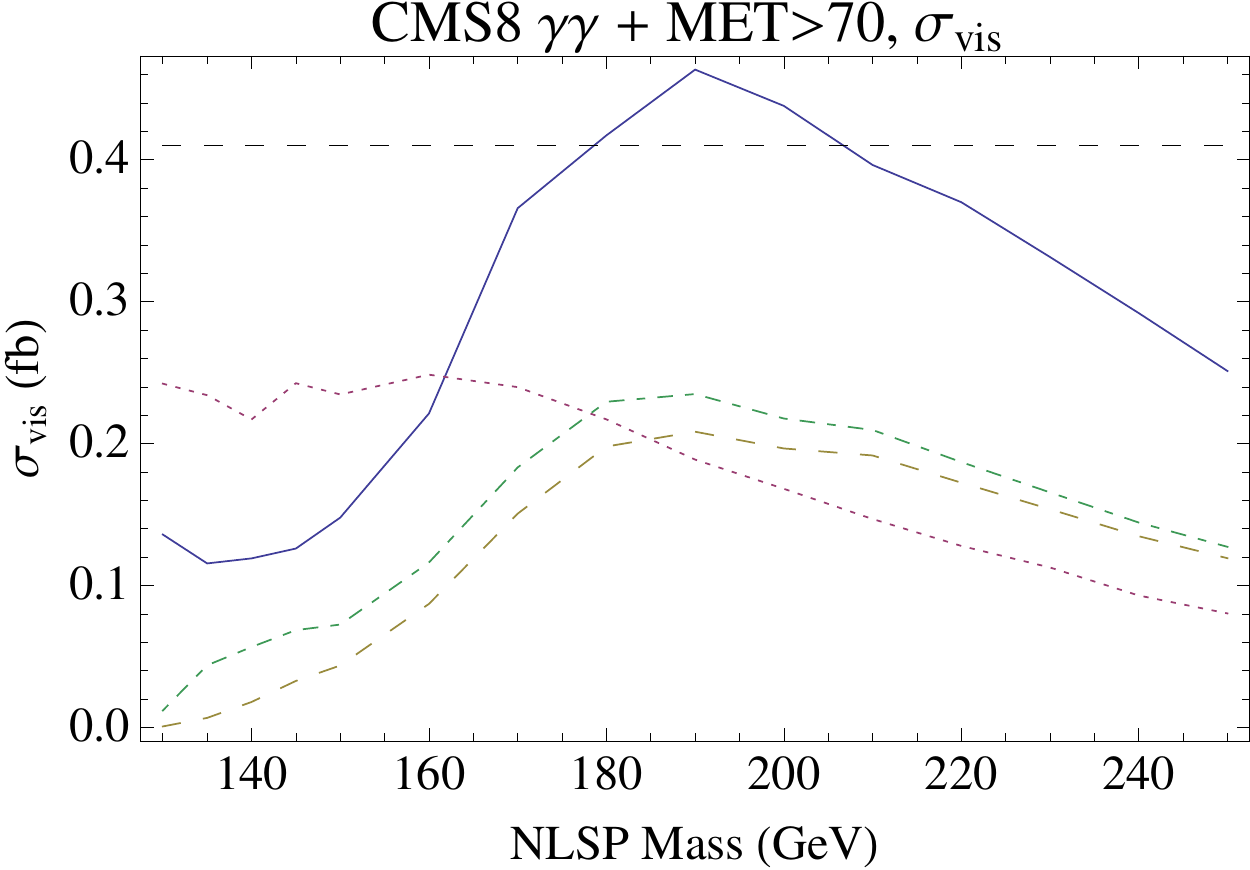}
                \label{fig:directDiphotonMET70}
		  }
               \hspace{5mm} 
 	\subfigure[]{
                \includegraphics[width=2.5in]{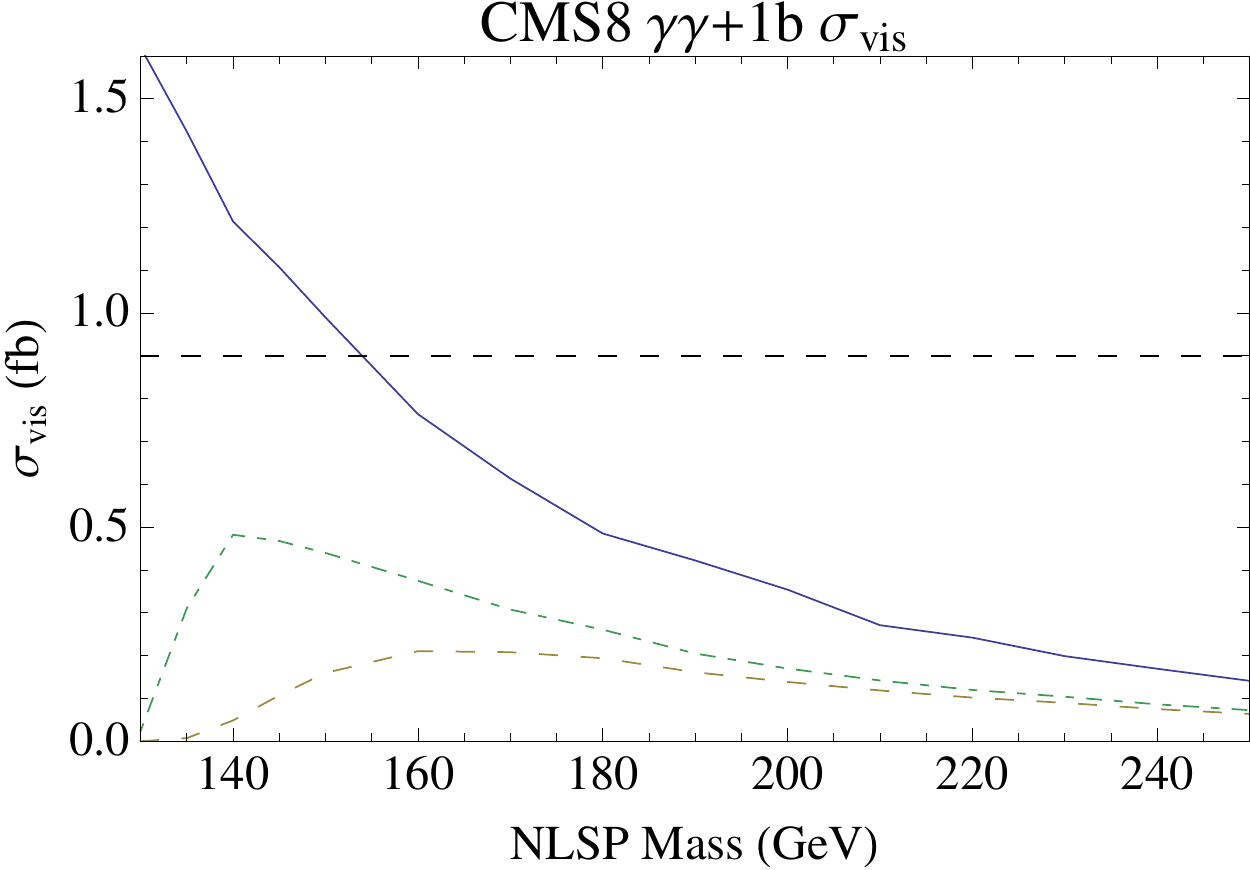}
                \label{fig:directDiphoton1b}
		  }

        \caption{Rates in diphoton channels for direct production. Solid is a mixed wino-bino NLSP, dotted is a wino co-NLSP, dashed is a Higgsino NLSP at $\tan\beta=1.5$, and dashed dotted is a Higgsino NLSP at $\tan\beta=1.1$. (a) Inclusive CMS8 diphoton rate relative to the SM Higgs rate. (b) Visible cross section in CMS8 $\gamma\gamma+\MET>70\GeV$ search. Dashed horizontal is the expected 95\% limit in $20\fb^{-1}$. (c) Visible cross section in $\gamma\gamma+1b$ search described in sec.~\ref{sec:ggb}. Dashed horizontal is the expected 95\% limit in $20\fb^{-1}$. \label{fig:DirectDiphoton}}
\end{figure}

For a mixed wino-bino NLSP or a higgsino-like NLSP, the charginos can promptly decay to the NLSP which then decays to a weakly coupled LSP, so the dominant signatures are in $(hh/hZ/ZZ) + \MET$. The Tevatron and LHC signatures for direct production of higgsinos decaying to gravitinos have been discussed in~\cite{Matchev:1999ft,ChiAtTeV,ChiAtLHC}, with emphasis on the situations where the branching ratio to $Z$ dominates or is comparable to the Higgs branching ratio. We focus instead on the case where the $hh$ mode is dominant, which in the higgsino case at low masses requires $\tan\beta\sim1$.  These modes are even less constrained than the wino co-NLSP because of their smaller branching fraction to leptonic final states. Refs.~\cite{Matchev:1999ft,ChiAtTeV} found that a dedicated Tevatron search in $hh\to4b+\MET$ could be sensitive to a higgsino NLSP with masses $\sim 150-200$ GeV, but to date the most constraining search has been a CDF search~\cite{Aaltonen:2009fm} for gluinos decaying to sbottoms requiring 2$b$-tags and MET $>$ 70 GeV, which does not constrain either case even assuming the same efficiencies as the targeted $\tilde g \to b\tilde b$ channel. Ref.~\cite{ChiAtLHC} briefly discussed the possibility of a $4b+\MET$ search at the LHC, but again we find that existing searches  (CMS $\alpha_T$ and ATLAS $3b+\MET$) do not set a limit on direct production.

The signals in the inclusive diphoton, diphoton+MET, and diphoton+1$b$ channels are shown in fig.~\ref{fig:DirectDiphoton} for the two wino NLSP possibilities assuming $100\%$ branching fractions for the NLSP to the Higgs, and for a higgsino NLSP decaying to a gravitino LSP at $\tan\beta = 1.1$ and $\tan\beta = 1.5$. Only the mixed wino-bino NLSP with prompt decays of the charginos to the NLSP has a potentially interesting signal in these channels, due to its large production cross section compared to a higgsino NLSP and the potential to produce two Higgses in each event, unlike the wino co-NLSP scenario. For masses lighter than 150 GeV it can have a signal of 20-30\% the SM rate in the inclusive $\gamma\gamma$ channel and can be potentially observed in the $\gamma\gamma+1b$ channel in 20$\fb^{-1}$. The rate in the $\gamma\gamma+2b$ channel is too low to afford sensitivity. The $\gamma\gamma+\MET$ search is just barely sensitive enough to excludes masses around $200\GeV$ in 20$\fb^{-1}$, but for lower masses the MET signal is too suppressed. Again, the rate in a lower MET bin can be several times higher and may increase the sensitivity for a lighter wino.  Although the rates for the other NLSP scenarios are too small to be observable from just direct production, in the context of the simplified squark production model the additional contribution from direct production can slightly enhance the overall sensitivity of the diphoton channels.

\section{Conclusions} \label{sec:Conclusions}

We have shown that in a variety of models the decays of a neutralino NLSP can give copious extra Higgs boson production while remaining consistent with constraints from BSM searches. This leads to potentially observable signatures in Higgs resonance searches. Well-motivated realizations of this scenario include wino- and bino-like NLSPs with a singlino/goldstino/string photino LSP, or a higgsino-like NLSP with a gravitino LSP. We have focused on pair production of light third generation squarks, motivated by naturalness of the electroweak scale. A wide class of realistic models can be well described by the simplified model of fig.~\ref{fig:SimpModel}, leading to final states with two bottom quarks, two Higgses, and missing energy. Existing searches constrain these models to have squark masses $\gtrsim 260\GeV$ and NLSP masses $\lesssim 200\GeV$, as summarized in fig.~\ref{fig:CombinedLimits}. For light squarks we argue that the most effective search strategy to probe these models is to target the $h \to \gamma \gamma$ decay mode, which would give a clear signal of a Higgs resonance. A search in $\gamma \gamma + 2b$ final states would have high signal efficiency and very low SM background, and could give high signal significance in 8 TeV LHC data for squarks up to $\sim 350 \GeV$ (figure ~\ref{fig:2bdiscovery}).   
    
Although a $\gamma \gamma + 2b$ search is expected to be the best probe for the models we consider, even existing analyses can have some sensitivity with further 8 TeV data. Of particular interest is the possibility that the BSM production rate is high enough to account for the high signal strength reported by ATLAS and CMS in the inclusive diphoton channels. This would require squarks near the limits placed by SUSY searches (see fig.~\ref{fig:ggInclusive}). Some excess are then expected for these searches in 8 TeV data, though this would be difficult to link to Higgs physics. More informative would be the searches for associated Higgs production, namely the $\gamma \gamma + \ell$ and $\gamma \gamma + \MET$ analyses, which would all have rates well in excess of the Standard Model prediction. As discussed in sec.~\ref{sec:OtherSearch}, little or no enhancement is expected in most searches in other Higgs decay modes such as $WW^*$ or $\tau \tau$. Therefore, observing excesses in only the diphoton channels and particularly the associated production channels would suggest our class of models. In the parameter space where observable excesses can occur in these channels, a search in $\gamma \gamma + 2b$ can likely give a $5\sigma$ discovery in 20$\fb^{-1}$.

Although we have focused on a particular SUSY simplified model in this work, similar phenomenology can arise in many other models. Final states very similar to those in our simplified model can arise for example from decays of new vector-like quarks to a Higgs and third generation quarks, especially in the decays of bottom-like quarks, e.g. $b' \to h b$. More generally, much of what we have said is relevant for pair production of any new particles that decay dominantly to Higgses. The diphoton decay mode still offers an excellent opportunity for identifying the Higgs resonance, and a $\gamma \gamma + 1b$ search can still have good efficiency because of the large Higgs branching ratio to $b \bar{b}$.

Finally, while this work has explored one well motivated scenario in which non-standard Higgs channels can become sensitive, we emphasize the importance of searching for Higgs production in as many channels as possible. Abandoning all theoretical prejudice, little is currently known about the production and decay of the newly observed resonance at 125 GeV, and there remains open the possibility of observing significant departures from Standard Model expectations in the very near future.

\section*{Acknowledgments}

We would like to thank Masha Baryakhtar, Timothy Cohen, Nathaniel Craig, Savas Dimopoulos, Peter Graham, Jo-Anne Hewett, Caitlin Malone, John March-Russell, Tom Rizzo, Surjeet Rajendran, Ken van Tilburg and Giovanni Villadoro for useful discussions. We especially thank Timothy Cohen and Peter Graham for giving comments on a draft of this work before publication. P.S. is supported by a Stanford Graduate Fellowship, and K.H. is supported by a NSF Graduate Research Fellowship under Grant No. DGE-0645962. This work was partially supported by ERC grant BSMOXFORD no. 228169.

\end{document}